\documentclass[11pt,preprint]{aastex}
\usepackage{amsmath}

\usepackage{color}
\usepackage[plainpages=false, colorlinks=true, anchorcolor=blue,
linkcolor=blue, citecolor=blue, bookmarks=false]{hyperref}

\usepackage{ulem}

\newcommand{\p}{\partial}
\newcommand{\pr}{\prime}
\newcommand\simgt{\lower.5ex\hbox{$\; \buildrel > \over \sim \;$}}
\newcommand\simlt{\lower.5ex\hbox{$\; \buildrel < \over \sim \;$}}

\newcommand{\pc}{\;\rm{pc}}
\newcommand{\cm}{\;\rm{cm}}

\newcommand{\mH}{m_{\rm{H}}}
\newcommand{\kB}{k_{\rm{B}}}
\newcommand{\Kel}{\;{\rm K}}
\newcommand{\eV}{\;{\rm eV}}
\newcommand{\condunit}{\;{\rm erg}\,{\rm cm}^{-1}\,{\rm s}^{-1}\,{\rm K}^{-1}}

\newcommand{\sigpi}{\sigma_{\rm{ph}}}
\newcommand{\vecJ}{\mathbf{J}}
\newcommand{\veco}{{\boldsymbol{\omega}}}
\newcommand{\jz}{j_z} 
\newcommand{\jzp}{j_z^{\pr}} 
\newcommand{\vz}{v_z}
\newcommand{\Bx}{B_x}
\newcommand{\vecv}{\boldsymbol{v}}
\newcommand{\vecB}{\boldsymbol{B}}
\newcommand{\hatn}{\hat{\mathbf{n}}} 
\newcommand{\hatx}{\hat{\mathbf{x}}}
\newcommand{\haty}{\hat{\mathbf{y}}}
\newcommand{\hatz}{\hat{\mathbf{z}}}

\newcommand{\rhou}{\rho_1}
\newcommand{\rhod}{\rho_2}
\newcommand{\Pu}{P_{1}} 
\newcommand{\Pd}{P_{2}}
\newcommand{\Fph}{F_{\rm ph}}
\newcommand{\vzu}{v_{z1}}
\newcommand{\vzd}{v_{z2}}
\newcommand{\Bxu}{B_{x1}}
\newcommand{\Bxd}{B_{x2}}

\newcommand{\gb}{G}
\newcommand{\alphah}{\hat{\alpha}}
\newcommand{\OD}{\Omega_{D}} 

\newcommand{\cs}{c_{\rm s}}
\newcommand{\csu}{c_{\rm s1}}
\newcommand{\csd}{c_{\rm s2}}
\newcommand{\vecvA}{\boldsymbol{v}_{\rm A}}
\newcommand{\vA}{v_{\rm A}}
\newcommand{\vAu}{v_{\rm A1}}

\newcommand{\MA}{\mathcal{M}_{\rm{A}}}
\newcommand{\MAu}{\mathcal{M}_{\rm{A1}}}
\newcommand{\MAd}{\mathcal{M}_{\rm{A2}}}
\newcommand{\Ms}{\mathcal{M}_{\rm{S}}}
\newcommand{\Msu}{\mathcal{M}_{\rm{S1}}}
\newcommand{\Msd}{\mathcal{M}_{\rm{S2}}}
\newcommand{\Mms}{\mathcal{M}_{\rm{M}}}
\newcommand{\Mmsu}{\mathcal{M}_{\rm{M}1}}
\newcommand{\Mmsd}{\mathcal{M}_{\rm{M}2}}
\newcommand{\MD}{\mathcal{M}_{\rm{D}}}
\newcommand{\MR}{\mathcal{M}_{\rm{R}}}

\newcommand{\betau}{\beta_{1}}
\newcommand{\betad}{\beta_{2}}
\newcommand{\Gammau}{\Gamma_{1}}
\newcommand{\Gammad}{\Gamma_{2}}

\newcommand{\nuf}{\nu_{\rm{f}}}

\newcommand{\nua}{\nu_{\rm{a}}}
\newcommand{\nuau}{\nu_{\rm{a1}}}
\newcommand{\nuad}{\nu_{\rm{a2}}}
\newcommand{\nuv}{\nu_{\rm{v}}}
\newcommand{\nus}{\nu_{\rm{s}}}

\newcommand{\nuAp}{\nu_{{\rm s2+}}}
\newcommand{\nuAm}{\nu_{{\rm s2-}}}
\newcommand{\nuApm}{\nu_{{\rm s2\pm}}}
\newcommand{\nuspm}{\nu_{{\rm s\pm}}}
\newcommand{\nuppm}{\nu_{{\rm p}\pm}}
\newcommand{\nupp}{\nu_{{\rm p}+}}
\newcommand{\nupm}{\nu_{{\rm p}-}}
\newcommand{\Deltaf}{\Delta}

\newcommand{\vecvp}{\boldsymbol{v}^{\pr}}
\newcommand{\vecBp}{\boldsymbol{B}^{\pr}}
\newcommand{\vecbp}{\boldsymbol{b}^{\pr}}

\newcommand{\veck}{\boldsymbol{k}}
\newcommand{\kx}{k_x}
\newcommand{\ky}{k_y}
\newcommand{\kz}{k_z}

\newcommand{\rhop}{\rho^{\pr}}
\newcommand{\rhopu}{\rho_1^{\pr}}

\newcommand{\Pp}{P^{\pr}}

\newcommand{\pp}{p^{\pr}}
\newcommand{\ppu}{p_1^{\pr}}
\newcommand{\ppd}{p_2^{\pr}}
\newcommand{\vxp}{v_{x}^{\pr}}
\newcommand{\vyp}{v_{y}^{\pr}}
\newcommand{\vzp}{v_{z}^{\pr}}
\newcommand{\vxpu}{v_{x1}^{\pr}}
\newcommand{\vypu}{v_{y1}^{\pr}}
\newcommand{\vzpu}{v_{z1}^{\pr}}
\newcommand{\vxpd}{v_{x2}^{\pr}}
\newcommand{\vypd}{v_{y2}^{\pr}}
\newcommand{\vzpd}{v_{z2}^{\pr}}
\newcommand{\bxp}{b_{x}^{\pr}}
\newcommand{\byp}{b_{y}^{\pr}}
\newcommand{\bzp}{b_{z}^{\pr}}
\newcommand{\bxpu}{b_{x1}^{\pr}}
\newcommand{\bypu}{b_{y1}^{\pr}}
\newcommand{\bzpu}{b_{z1}^{\pr}}
\newcommand{\bxpd}{b_{x2}^{\pr}}
\newcommand{\bypd}{b_{y2}^{\pr}}
\newcommand{\bzpd}{b_{z2}^{\pr}}
\newcommand{\Bxp}{B_{x}^{\pr}}
\newcommand{\Byp}{B_{y}^{\pr}}
\newcommand{\Bzp}{B_{z}^{\pr}}

\shorttitle{Instability of Magnetized Ionization Fronts} %
\shortauthors{Kim \& Kim}

\begin{document}

\title{Instability of Magnetized Ionization Fronts Surrounding
  \ion{H}{2} Regions}

\author{Jeong-Gyu Kim and Woong-Tae Kim}

\affil{Center for the Exploration of the Origin of the Universe
  (CEOU), Astronomy Program, Department of Physics \& Astronomy,\\
  Seoul National University, Seoul 151-742, Republic of Korea}
\email{jgkim@astro.snu.ac.kr, wkim@astro.snu.ac.kr}

\slugcomment{ApJ accepted}

\begin{abstract}
An ionization front (IF) surrounding an \ion{H}{2} region is a sharp
interface where a cold neutral gas makes transition to a warm ionized
phase by absorbing UV photons from central stars. We investigate the
instability of a plane-parallel D-type IF threaded by parallel
magnetic fields, by neglecting the effects of recombination within the
ionized gas. We find that weak D-type IFs always have the post-IF
magnetosonic Mach number $\mathcal{M}_{\rm M2} \leq 1$. For such
fronts, magnetic fields increase the maximum propagation speed of the
IFs, while reducing the expansion factor $\alpha$ by a factor of
$1+1/(2\beta_1)$ compared to the unmagnetized case, with $\beta_1$
denoting the plasma beta in the pre-IF region. IFs become unstable to
distortional perturbations due to gas expansion across the fronts,
exactly analogous to the Darrieus-Landau instability of ablation
fronts in terrestrial flames. The growth rate of the IF instability is
proportional linearly to the perturbation wavenumber as well as the
upstream flow speed, and approximately to $\alpha^{1/2}$. The IF
instability is stabilized by gas compressibility and becomes
completely quenched when the front is D-critical. The instability is
also stabilized by magnetic pressure when the perturbations propagate
in the direction perpendicular to the fields. When the perturbations
propagate in the direction parallel to the fields, on the other hand,
it is magnetic tension that reduces the growth rate, completely
suppressing the instability when $\mathcal{M}_{\rm M2}^2 < 2/(\beta_1
- 1)$. When the front experiences an acceleration, the IF instability
cooperates with the Rayleigh-Taylor instability to make the front more
unstable.
\end{abstract}


\section{Introduction}

\ion{H}{2} regions are volumes of ionized gas formed by absorbing UV
photons emitted by newborn massive stars. Since the ionized gas is
overpressurized by about two orders of magnitude compared to the
surrounding neutral medium, it naturally expands to affect the
structure and dynamics of a surrounding interstellar medium (ISM).
Photoionization appears to play a dual role in regulating star
formation in the ISM.  On one hand, it can evaporate and disrupt
parental molecular clouds, limiting the efficiency of star formation to
about a few percents \citep{mat02,kru06,wal12,dal12}. On the other
hand, the expansion of \ion{H}{2} regions may sustain turbulence in
clouds \citep{mel06,kru06,gri10,dal12}, and trigger gravitational
collapse of compressed shells \citep{elm77,hos06,dal09,iwa12} as well
as pre-existing clumps in the surrounding medium
\citep[e.g.,][]{san82,ber89,bis11}, tending to promote further star
formation. Understanding how \ion{H}{2} regions evolve may thus be the
first step to understand the effect of the star formation feedback on
the ISM (see \citealt{kru14}, for a recent review).

A number of pioneering studies have explored the dynamical expansion of
an \ion{H}{2} region (\citealt{str39,kah54,gol58,axf61,mat65}; see also
\citealt{yor86,shu92}). Soon after a central ionizing source is turned
on, an ionization front (IF) develops to separate a warm ionized gas
with temperature $T \sim 10^4 \Kel$ from a cold neutral gas with $T
\sim 10^2 \Kel$, with thickness of order of only a few photon mean free
paths. At early times, the ionizing photon flux is very large and the
IF advances into the neutral medium supersonically, without inducing
gas motions.  In this early phase, the IF is termed ``weak R-type''
\citep{kah54}.  After roughly a few recombination times (typically
$\sim 10^3$ yrs), the initial Str\"{o}mgren sphere is established
within which the recombination rate balances the ionizing rate. At this
point, the IF stops propagating and turns to a ``R-critical'' front.
Since the ionized gas behind the IF moves at the sonic speed with
respect to the IF, it is able to launch shock waves into the regions
ahead of the IF, which in turn makes the R-critical front switch to a
``D-critical'' front. After this transition, the expansion of the IF is
driven by the pressure difference between the ionized and neutral gas.
The IF becomes ``weak D-type'' and moves subsonically with respect to
the neutral gas. Given that the main-sequence lifetime of O/B stars is
typically $\sim10^6-10^7$ yrs, \ion{H}{2} regions spend most of their
lives in the weak D-type phase.

While the early efforts based on one-dimensional models under spherical
symmetry provide valuable insights on the overall evolution, \ion{H}{2}
regions are abound with various substructures such as globules,
filaments, gaseous pillars (or ``elephant trunks''), etc. that cannot
be explained by the one-dimensional models (e.g.,
\citealt{sug91,hes96,chu06}). One promising explanation may be that
these non-spherical structures result from pre-existing density
inhomogeneities and/or turbulent motions in the background medium.
Numerical simulations indeed showed that ionizing radiation
illuminating on a turbulent, inhomogeneous medium forms
non-axisymmetric structures, elongated away from the ionizing source,
which may undergo gravitational collapse (e.g.,
\citealt{gri10,art11,tre12a,tre12b}).

Although less well-recognized, instability of IFs can be another route
to the formation of non-spherical structures. \citet{fri54} and
\citet{spi54} suggested that IFs are susceptible to the Rayleigh-Taylor
instability (RTI) when the front is accelerating away from the central
source, which may occur due to a steep gradient of the background
density or time-varying radiation intensity. \citet{van62} performed a
linear stability analysis of an unmagnetized, planar weak D-type IF by
including a steady motion of gas relative to the front. He found that
such steady-state IFs even without acceleration are unstable at all
wavelengths, with the growth rate inversely proportional to the
perturbation wavelength. \citet{axf64} subsequently showed that the
long-wavelength modes are stabilized by the attenuation of radiation
due to hydrogen recombination in the perturbed ionized gas if
the radiation is normal to the front. Allowing for finite temperature
ratio between the neutral and ionized phases, \citet{saa66} obtained an
approximate, closed-form expression for the growth rate of the IF
instability. Later, \citet{sys97} discovered the existence of
long-wavelength unstable modes even in the presence of recombination.
\citet{wil02} further extended the previous studies by including
oblique incident angles of the ionizing radiation relative to the
front, and carried out local isothermal simulations of the instability
to study its nonlinear development. By running simulations of expanding
\ion{H}{2} regions, \citet{gar96} and \citet{wha08} suggested that a
shocked shell undergoes a thin-shell instability \citep{vis83} to grow
into sword-like structures. \cite{ric14} performed a stability analysis
of accelerating IFs in the limit of incompressible fluids and showed
that the recombination stabilizes RTI of IFs (see also
\citealt{miz05,par14}).

Despite these efforts on the IF instability, however, there are still
two main issues that remain to be answered.  First, what is the
physical nature of the instability? \citet{van62} argued that the IF
instability results from a mechanism similar to the ``rocket effect''
of gradually evaporating clouds \citep{kah54,oor55}. However, the
rocket effect acting on the IF can always make the IF move away from
the central source and is thus unable to explain the wavy movement of
the perturbed IF toward the source, which is the key in the operation
of the IF instability, as we will show below.  The second issue
concerns the effect of magnetic fields on the IF instability.
Observations indicate that the ISM around \ion{H}{2} regions are
permeated by magnetic fields. The typical values for the ratio of
magnetic to thermal pressure are $\sim 0.3$ and $\sim 0.04$ in cold
neutral and molecular clouds, respectively \citep{cru99,hei05}. The
line-of-sight component of magnetic fields {\it inside} five Galactic
\ion{H}{2} regions based on Faraday rotation diagnostics is estimated
to be in the range of $B_{\rm los} \sim 2$--$6\,\mu{\rm G}$,
corresponding to subthermal magnetic pressure (\citealt{har11}; see
also \citealt{hei80,hei81,rod12}). On the other hand, Zeeman
observations of \ion{H}{1} and OH absorption lines reveal that the
interfaces between the ionized and molecular gases are strongly
magnetized with $B_{\rm los} \sim 50$--$300\,\mu{\rm G}$ and $\sim
250$--$750\,\mu{\rm
  G}$ for Orion's Bar and M17, respectively
\citep{bro99,bro01,bro05}. Using hydrostatic models,
\citet{pel07,pel09} confirmed that these clouds are indeed dominated
by magnetic pressure.

Importance of magnetic fields associated with IFs has been emphasized
by several authors.  For instance, \citet{las66} showed that D-type IFs
may contain strong magnetic fields when preceded by an isothermal
shock. \citet{red98} calculated the jump conditions for IFs with
magnetic fields parallel to the fronts. \citet{wil00} show that an IF
with oblique magnetic fields drives fast- and slow-mode shocks
separately as it slows down, suggesting IFs should be subclassified
according to the propagation speed. Numerical simulations showed that
\ion{H}{2} regions threaded by uniform magnetic fields expand
aspherically and that both shock strength and density contrast across
the IF are reduced in regions where the magnetic fields are parallel to
the front \citep{kru07,mac11}. Although these simulations also reported
the deformation of IFs during nonlinear evolution, they lacked spatial
resolution to capture the instability of IFs.

In this paper we address the two issues mentioned above by performing
a linear stability analysis of weak D-type IFs with and without
magnetic fields. We also include the acceleration/deceleration term in
the momentum equation in order to study the combined effects of the
RTI and IF instability. We will show that the operating mechanism
behind the IF instability is the same as that of the Darrieus-Landau
instability (DLI) found in terrestrial flames
\citep[e.g.][]{lan59,zel85}. The DLI is inherent to any evaporative
interfacial layer through which a cold dense gas expands to become a
warm rarefied gas by absorbing heat.  Examples include carbon
deflagration fronts in Type Ia supernovae \citep{bel04,dur04},
evaporation fronts in the multi-phase ISM \citep{ino06,sto09,kim13},
and ablation fronts in inertial confinement fusion
\citep{byc08,mod09}. We will also show that magnetic fields stabilize
the IF, although the roles of magnetic pressure and tension are
different depending on the propagation direction of perturbations. We
will further show that the RTI can enhance the IF instability when the
IF is accelerating away from the ionizing source, while buoyancy
stabilizes large-scale modes for decelerating IFs.

The rest of this paper is organized as follows. In Section \ref{s:2},
we present the steady-state equilibrium solutions of IFs, and provide
approximate expressions for the density jumps for weak D-type,
D-critical, and R-critical fronts.  In Section \ref{s:3}, we describe
our method of a linear stability analysis by classifying the basis
modes of perturbations and presenting the perturbed jump conditions
across an IF. In Section \ref{s:hd}, we revisit the case of
unmagnetized IFs and clarify the physical nature of the IF instability.
In Section \ref{s:mhd}, we analyze the stability of magnetized IFs for
perturbations that propagate along the direction either perpendicular
(Section \ref{s:mhd1}) or parallel (Section \ref{s:mhd2}) to the
magnetic fields. The growth rate of the instability is presented for
both incompressible and compressible cases. In Section \ref{s:sum}, we
summarize our findings and discuss their astrophysical implications.

\section{Steady Ionization Fronts}\label{s:2}

\subsection{Basic Equations}

In this paper we investigate a linear stability of a magnetized D-type
IF subject to an effective external gravity.  We treat the IF as a
plane-parallel surface of discontinuity and do not include the effects
of heat conduction, magnetic diffusion, and gaseous self-gravity. The
basic equations of ideal magnetohydrodynamics (MHD) are
\begin{equation}\label{e:con}
  \dfrac{\p \rho}{\p t} + \nabla \cdot (\rho \vecv) = 0 \,,
\end{equation}
\begin{equation}\label{e:mom}
  \dfrac{\p \vecv}{\p t} + (\vecv \cdot \nabla)\vecv =
  -\dfrac{1}{\rho}\nabla
  \left( P + \dfrac{|\vecB|^2}{8\pi} \right) -
  \dfrac{1}{4\pi\rho}(\vecB\cdot\nabla)\vecB +
\mathbf{g}\,,
\end{equation}
\begin{equation}\label{e:ind}
  \dfrac{\p\vecB}{\p t} = \nabla\times(\vecv\times\vecB)\,,
\end{equation}
and
\begin{equation}\label{e:div}
  \nabla \cdot \vecB = 0\,,
\end{equation}
where $\rho$, $P$, $\vecv$, and $\vecB$ denote the gas density, thermal
pressure, velocity, and magnetic fields, respectively. The constant
acceleration $\mathbf{g}$ in Equation (\ref{e:mom}) is to represent a
situation where the IF propagation away from a central source speeds up
or slows down, which may occur due to a nonuniform background density
and/or the geometrical dilution of UV radiation intensity. Since the
thermal time scale is usually very short compared to the dynamical time
scale of \ion{H}{2} regions, we adopt an isothermal equation of state
\begin{equation}\label{e:eos}
  P=\cs^2\rho\,,
\end{equation}
where $\cs$ is the speed of sound that takes different values in the
regions ahead and behind of an IF separating ionized and neutral
gases.

The mass flux across an IF is determined by the amount of UV photons
irradiated to it. Let $\vecJ(\boldsymbol{\omega})$ denote the photon
number intensity at the front in the direction $\veco \equiv
\vecJ/|\vecJ|$ \citep{van62}. Then, the photon flux at the front is
equal to $\Fph \equiv - \int \hatn \cdot \vecJ d\omega$, where $\hatn$
is the unit vector normal to the front and the integration is taken
over the solid angle of the hemisphere directed toward the ionizing
source. In this work, we consider only the case in which the incident
ionizing radiation is the normal to the front, i.e., $J(\veco) = \Fph
\delta(\veco + \hatn)$, with $\delta$ being the Dirac delta
function.\footnote{When the incident radiation is oblique to the front
  normal, the IF becomes overstable rather than unstable due to the
  phase differences between the front deformation and density
  perturbations \citep{van62,wil02}.} The condition that all arriving
photons are consumed at the IF can then be expressed as
\begin{equation}\label{e:pho}
  \rho\vecv \cdot \hatn = \mH \Fph\,,
\end{equation}
where $\mH$ is the mass per particle in the neutral atomic medium.

\subsection{Steady-State Configurations}

As an undisturbed state, we seek for one-dimensional steady-state
solutions of Equations (\ref{e:con})--(\ref{e:pho}).  The properties of
magnetized IFs were presented by \citet{red98}.  Here, we focus on
finding the background configurations of IFs, the stability of which
will be explored later. We also present approximate solutions of
magnetized IFs for critical and weak D-type fronts.

We place an IF at $z=0$, with $\hatz=\hatn$.  Seen in the stationary
IF frame, a cold neutral gas located at $z<0$ is moving toward the
positive-$z$ direction, and becomes ionized and heated upon crossing
the IF by absorbing UV photons. The ionized gas in the region at $z>0$
flows away from the IF with speed and density different from those of
the neutral gas. For simplicity, we assume that the initial magnetic
fields are oriented along the $x$-direction such that $\vecB =
\Bx\hatx$, parallel to the front.  We further assume that the effect
of $\mathbf{g}=g\hatz$ is negligible in the initial configurations.
Equations (\ref{e:con})--(\ref{e:pho}) are then combined to give
\begin{equation}\label{e:con_st}
  \jz \equiv  \rhou\vzu = \rhod\vzd = \mH \Fph\,,
\end{equation}
\begin{equation}\label{e:mom_st}
  \Pu + \rhou\vzu^2 + \dfrac{\Bxu^2}{8\pi} = \Pd + \rhod\vzd^2 +
  \dfrac{\Bxd^2}{8\pi}\,,
\end{equation}
\begin{equation}\label{e:ind_st}
  \Bxu\vzu = \Bxd\vzd\,.
\end{equation}
Here and hereafter, we use the subscripts ``1'' and ``2'' to indicate
physical quantities evaluated at the neutral-gas region (at $z<0$) and
the ionized-gas region (at $z>0$), respectively. In Equation
\eqref{e:con_st}, $\jz$ denotes the (constant) mass flux across the IF.

We define the dimensionless expansion factor
\begin{equation}\label{e:mu}
  \alpha \equiv \frac{\vzd}{\vzu} = \frac{\rhou}{\rhod} =
  \frac{\Bxu}{\Bxd}\,,
\end{equation}
and the heating factor
\begin{equation}\label{e:theta}
  \theta \equiv \frac{\csd^2}{\csu^2}\,.
\end{equation}
The heating factor across the IF approximately equals $2T_2/T_1$,
where $T_1$ and $T_2$ are the temperatures of the neutral and ionized
gases, respectively.  Since $T_1=10^2$ K and $T_2=10^4$ K typically,
we in this work take a fiducial value of $\theta = 200$.  We define
the plasma parameter as
\begin{equation}
 \beta \equiv \cs^2/\vA^2\,,
\end{equation}
with $\vA \equiv \Bx/\sqrt{4\pi\rho}$ being the Alfv\'{e}n speed.  We
also define the sonic Mach number, the Alfv\'enic Mach number, and the
magnetosonic Mach number as
\begin{equation}\label{e:defM}
 \Ms \equiv \frac{\vz}{\cs}\,, \;\;\;
 \MA \equiv \frac{\vz}{\vA}\,, \;\;\;
 \Mms \equiv \frac{\vz}{(\cs^2+\vA^2)^{1/2}}\,,
\end{equation}
respectively. It then follows that $\Mms^2=\Ms^2/(1+\beta^{-1}) =
\MA^2/(1+\beta)$.

Using Equations \eqref{e:mu}, \eqref{e:theta}, and \eqref{e:defM}, one
can show that
\begin{equation}\label{e:betad}
  \betad = \alpha\theta\betau\,,
\end{equation}
\begin{equation}\label{e:MAad}
  \MAd^2/\MAu^2=\alpha^3\,,
\end{equation}
and
\begin{equation}\label{e:MMad}
 \frac{\Mmsd^2}{\Mmsu^2} = \alpha^3 \frac{1+\betau^{-1}}{\theta\alpha+\betau^{-1}}\,.
\end{equation}
We combine Equations \eqref{e:con_st}--\eqref{e:ind_st} to eliminate
$\vzd$ and $\Bxd$ in favor of $\rhod$ and write the resulting equation
in dimensionless form as
\begin{equation}\label{e:fmu}
  \Msu^2\alpha^3 -
  \left(1 + \Msu^2 + \frac{1}{2\betau} \right)\alpha^2 +
  \theta\alpha + \frac{1}{2\betau} = 0\,,
\end{equation}
which is a cubic equation in $\alpha$ (\citealt{red98}; see also
\citealt{dra11}).

\subsubsection{Unmagnetized IFs}

For unmagnetized fronts ($\betau\rightarrow \infty$), Equation
(\ref{e:fmu}) is reduced to the following quadratic equation:
\begin{equation}\label{e:hd0}
 \Msu^2\alpha^2 - (\Msu^2+1) \alpha + \theta =0\,.
\end{equation}
Equation \eqref{e:hd0} has real solutions only if $\Msu\leq
\mathcal{M}_{\rm D}$ or $\Msu\geq \mathcal{M}_{\rm R}$, where
\begin{equation}\label{e:crM}
\begin{split}
  \mathcal{M}_{\rm D} & = \sqrt{\theta} - \sqrt{\theta -1} \approx
  1/(2\sqrt{\theta}), \\
  \mathcal{M}_{\rm R} & = \sqrt{\theta} + \sqrt{\theta -1} \approx
  2\sqrt{\theta},
\end{split}
\end{equation}
denote the sonic Mach numbers of the D- and R-critical fronts,
respectively \citep[e.g.,][]{kah54,spi78,shu92}. The corresponding
expansion factors are
\begin{equation}\label{e:crA}
\begin{split}
  \alpha_{\rm D} &= \theta + \sqrt{\theta^2 -\theta} \approx 2\theta\,, \\
  \alpha_{\rm R} &= \theta - \sqrt{\theta^2 - \theta} \approx 1/2\,,
\end{split}
\end{equation}
for the D- and R-critical fronts, respectively. Both D- and R-critical
IFs satisfy
\begin{equation}
 \mathcal{M}_{\rm S2,D} = \mathcal{M}_{\rm S2,R} = 1\,,
\end{equation}
exactly in the downstream side.

Fronts with $\Msu < \mathcal{M}_{\rm D}$ and $\Msu> \mathcal{M}_{\rm
  R}$ are called D-type and R-type IFs, respectively. For these,
Equation \eqref{e:hd0} has two real solutions for $\alpha$. IFs with a
smaller and larger density jump (i.e., smaller and larger $|\alpha-1|$)
are further termed ``weak'' and ``strong'' fronts, respectively. The
fact that the inflow velocity $\Msu$ relative to an IF in a steady
state cannot be arbitrary is because the temperature of the post-IF
region is prespecified to $\theta$ greater than unity. In the limit of
$\theta\rightarrow1$, $\mathcal{M}_{\rm D} = \mathcal{M}_{\rm R} = 1$,
and Equation (\ref{e:hd0}) yields $\alpha=1$ for a strong front and
$\alpha=\Msu^{-2}$ for a weak front, the second of which is simply the
jump condition for an isothermal shock.

\subsubsection{Magnetized IFs}

The presence of magnetic fields certainly changes the critical Mach
numbers as well as the expansion factors. Since the coefficients of the
third- and zeroth-degree terms are real and positive, Equation
\eqref{e:fmu} always has a negative real root. The other two roots
should thus be real and positive for physically meaningful fronts,
which limits the ranges of $\Msu$ for magnetized IFs. \citet{red98}
showed that the critical Mach number $\mathcal{M}_{\rm S1, crit}$ and
$\betau$ are related to each other via $\alpha$ by
\begin{equation}\label{e:Redman1}
 \mathcal{M}_{\rm S1,crit}^2 =
 \frac{\theta(\alpha^2+1)-2\alpha}{\alpha(\alpha-1)^2(\alpha+2)}\,,
\end{equation}
and
\begin{equation}\label{e:Redman2}
 \betau =\frac{(\alpha-1)^2(\alpha+2)}{2\alpha(2\theta\alpha-\alpha^2-\theta)}\,.
\end{equation}
The smaller and larger values of $\mathcal{M}_{\rm S1,crit}$
correspond to $\mathcal{M}_{\rm D}$ and $\mathcal{M}_{\rm R}$,
respectively, for magnetized fronts. Equations \eqref{e:Redman1} and
\eqref{e:Redman2} are combined to give
\begin{equation}\label{e:S2c}
  \mathcal{M}_{\rm S1,crit}^2
  = \frac{\theta}{\alpha^2} \left[1 + \frac{1}{\alpha\theta\betau}\right]\,,
\end{equation}
which yields
\begin{equation}
 \mathcal{M}_{\rm M2,crit}^2 = \frac{\mathcal{M}_{\rm S2,crit}^2}{1 +
 \betad^{-1}}= 1\,,
\end{equation}
exactly.  Therefore, weak D-type IFs always have $0< \Mmsd \leq 1$,
while $\Mmsd\geq 1$ for strong D-type IFs.  Substituting Equation
\eqref{e:S2c} into Equation \eqref{e:fmu}, we have
\begin{equation}\label{e:fmu2}
  \left(1 + \frac{1}{2\betau}\right)\alpha^3 - 2\theta \alpha^2
  +
  \left( \theta - \frac{3}{2\betau}\right)\alpha + \frac{1}{\betau} = 0,
\end{equation}
whose two positive roots correspond to the expansion factors for the
magnetized D- and R-critical fronts.  Note that Equation \eqref{e:fmu2}
reduces to Equation \eqref{e:crA} for $\betau\rightarrow\infty$.

\begin{figure}
\epsscale{0.55}\plotone{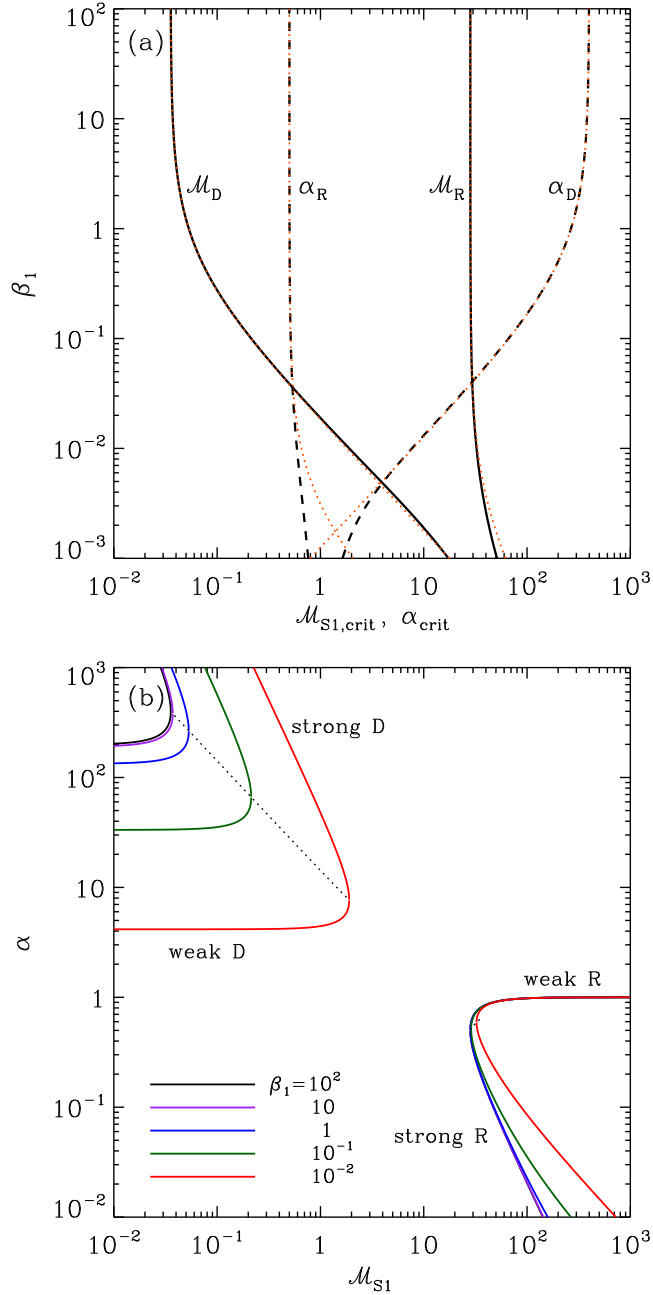} \caption{(a) Dependence on $\betau$
  of the critical Mach numbers $\mathcal{M}_{\rm D}$ and $\mathcal{M}_{\rm
    R}$ (black solid lines) and the critical expansion factors
  $\alpha_{\rm D}$ and $\alpha_{\rm R}$ (black dashed lines)
  for $\theta =200$. The red dotted lines give the
  analytic approximations (Equations \eqref{e:crM2} and
  \eqref{e:crA2}). (b) Expansion factors $\alpha$ of magnetic IFs with $\theta
  = 200$ for various $\betau$. The dotted curves mark the loci of the
  critical fronts where strong and weak solutions merge.}\label{f:01}
\end{figure}

Figure \ref{f:01}(a) plots the relationships between $\betau$ and
$\mathcal{M}_{\rm S1,crit}$ (black solid lines) and between $\betau$
and $\alpha_{\rm crit}$ (black dashed lines) for $\theta = 200$.
Clearly, $\mathcal{M}_{\rm D}$ and $\mathcal{M}_{\rm R}$ are close to
the unmagnetized value given in Equation (\ref{e:crM}) for
$\betau\simgt 10$. They increase as $\betau$ decreases.  For
$\theta\gg1$ and $\betau\theta\gg1$, one can expand Equations
\eqref{e:fmu2} relative to the unmagnetized solutions to show that the
critical expansion factors are approximately
\begin{equation}\label{e:crM2}
\begin{split}
  \alpha_{\rm D} & \approx 2\theta \left(1 + \frac{1}{2\betau}\right)^{-1}\,, \\
  \alpha_{\rm R} & \approx \frac{1}{2} +  \frac{5}{16\theta\betau}\,.
\end{split}
\end{equation}
The corresponding critical Mach numbers are
\begin{equation}\label{e:crA2}
\begin{split}
   \mathcal{M}_{\rm D}^2 & \approx \frac{1}{4\theta}\left(1+ \frac{1}{2\betau}\right)^2, \\
   \mathcal{M}_{\rm R}^2 & \approx 4\theta\left(1 + \frac{3}{4\betau\theta}\right).
\end{split}
\end{equation}
These are plotted as red dotted lines in Figure \ref{f:01}(a), in good
agreement with the full solutions for $\betau\simgt
10^{-2}$.\footnote{\citet{dra11} derived approximate expressions for
  the critical Mach numbers of magnetized IFs by taking $\alpha\gg1$
  and thus $\Bxu/\Bxd \gg1$ in Equation (\ref{e:mom_st}). In the limit
  of $\theta\gg1$ and $\betau\theta\gg1$, his results are equal to
  ours only for $\MD$. Since R-type IFs have $\alpha<1$ and $\Bxu <
  \Bxd$, the approximation he made is not valid for $\MR$.}

\begin{figure}
  \epsscale{0.55}\plotone{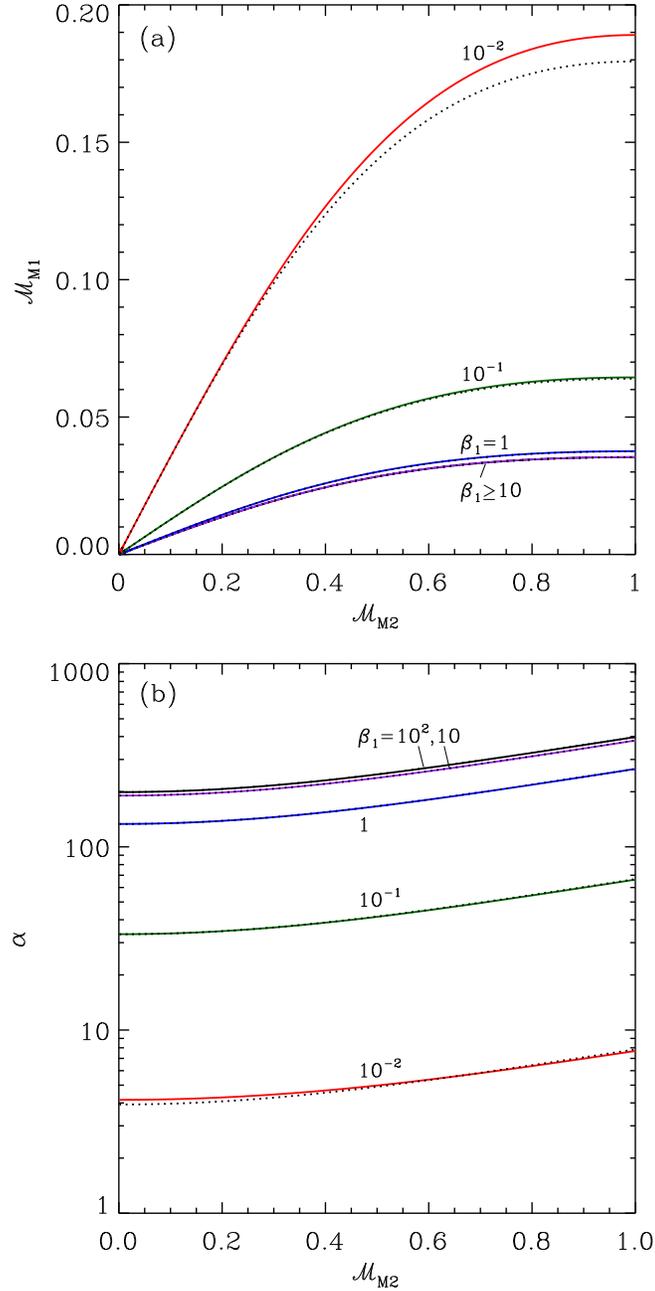}
  \caption{Dependence of (a) the upstream magnetosonic Mach number
    $\Mmsu$ and (b) the expansion factor $\alpha$ on the downstream
    magnetosonic Mach number $\Mmsd$ for weak D-type IFs with $\theta
    = 200$.  The solid lines give the full numerical results, while
    the dotted lines draw the approximate solutions (Equations
    \eqref{e:mu_app} and \eqref{e:Mmsu_app}).}\label{f:02}
\end{figure}

Figure \ref{f:01}(b) plots the expansion factors of magnetized IFs for
different values of $\betau$ as functions of $\Msu$ for $\theta=200$.
The dotted lines draw the loci of the D- and R-critical IFs.  Weak
R-type IFs have $\alpha\approx1$, and their physical conditions are
almost unaffected by the presence of magnetic fields \citep{las66}. For
weak D-type IFs displayed in the lower parts of the left curves,
however, magnetic fields lower the expansion factor considerably,
especially for $\betau \lesssim 1$.

Figure \ref{f:02} plots $\Mmsu$ and $\alpha$ as functions of $\Mmsd$
for weak D-type IFs with $\theta=200$. Magnetic fields clearly reduce
$\alpha$, while increasing $\Mmsu$ for given $\Mmsd$. Since $\Msu^2 \ll
1$ and $\betad \gg 1$ for weak D-type IFs, Equation \eqref{e:mom_st}
approximately gives
\begin{equation}\label{e:mu_app}
  \alpha \approx \theta \dfrac{1+\Msd^2}{1 +
  1/(2\betau)} \approx \theta \dfrac{1+\Mmsd^2}{1 +
  1/(2\betau)}\,,
\end{equation}
which, combined with Equation \eqref{e:MMad}, yields
\begin{equation}\label{e:Mmsu_app}
  \Mmsu \approx \frac{1}{\theta^{1/2}}
    \dfrac{1+1/(2\betau)}{(1 + \betau^{-1})^{1/2}}\dfrac{\Mmsd}{1
    +\Mmsd^2}\,,
\end{equation}
These are plotted in Figure \ref{f:02} as dotted lines, in good
agreement with the full solutions (better than 6\% for $\betau\geq
10^{-2}$). Note that the second approximation in Equation
\eqref{e:mu_app} follows from the approximation $\betad\gg1$, which is
usually the case unless $\betau\theta \lesssim 1$ (see Equation
\eqref{e:betad} and Figure \ref{f:02}). Equation \eqref{e:mu_app} then
indicates that the magnetic fields reduce the expansion factor $\alpha$
by a factor of $1+1/(2\betau)$ compared to the unmagnetized value.

\section{Perturbation Equations}\label{s:3}

We now apply small-amplitude perturbations to a steady IF with
$\vecvA=\vA\hatx$ found in the preceding section, and then explore
their stability. Since the background flow is uniform except at
$z=0$, Equations (\ref{e:con})--(\ref{e:eos}) can be linearized on
both sides of the front as
\begin{equation}\label{e:con_p}
  \left(\dfrac{\p}{\p t} + \vecv\cdot\nabla\right)\rhop = -
  \rho\nabla\cdot\vecvp\,,
\end{equation}
\begin{equation}\label{e:mom_p}
  \left(\dfrac{\p}{\p t} + \vecv\cdot\nabla\right)\vecvp =
  -\dfrac{1}{\rho}\nabla \left(
    \Pp + \dfrac{\vecB\cdot\vecBp}{4\pi} \right) +
  \dfrac{1}{4\pi\rho}(\vecB\cdot\nabla)\vecBp\,,
\end{equation}
\begin{equation}\label{e:ind_p}
  \left(\dfrac{\p}{\p t} + \vecv\cdot\nabla\right)\vecBp =
  (\vecB\cdot\nabla)\vecvp - \vecB(\nabla\cdot\vecvp)\,,
\end{equation}
\begin{equation}\label{e:div_p}
  \nabla \cdot \vecBp = 0\,,
\end{equation}
\begin{equation}\label{e:eos_p}
  \Pp =\cs^2 \rhop\,,
\end{equation}
where the primes denote the perturbed quantities. We assume that the
perturbations vary in space and time as
\begin{equation}\label{e:fourier}
  \propto \exp(i\veck\cdot \boldsymbol{x} + \Omega t )\,,
\end{equation}
where $\veck=(\kx, \ky, \kz)$ and $\Omega$ are the wavenumber and the
frequency of the perturbations, respectively. We take a convention
that $\kx$ and $\ky$ are real, while $\Omega$ and $\kz$ are complex.

Equations (\ref{e:con_p})--(\ref{e:eos_p}) are then reduced to
\begin{equation}\label{e:con_p2}
  \OD \rhop = -\rho(i\veck\cdot\vecvp)\,,
\end{equation}
\begin{equation}\label{e:mom_p2}
  \OD \vecvp = -i\veck \left(\frac{\cs^2}{\rho}\rhop + \vA \bxp
  \right) + i \kx\vA \vecbp\,,
\end{equation}
\begin{equation}\label{e:ind_p2}
  \OD \vecbp = i \kx\vA\vecvp - i \vA (\veck\cdot\vecvp)\hatx\,,
\end{equation}
\begin{equation}\label{e:div_p2}
  \veck\cdot\vecbp = 0\,,
\end{equation}
where $\OD \equiv \Omega + i\kz\vz$ is the Doppler-shifted frequency
and $\vecbp \equiv \vecBp/\sqrt{4\pi\rho}$ has the units of velocity.
Equations \eqref{e:con_p2}--\eqref{e:div_p2} are combined to yield
\begin{equation}\label{e:fdisp}
\begin{split}
  (\OD^2 + \kx^2\vA^2)\vecvp
   + [ (\cs^2+\vA^2) (\veck\cdot\vecvp) -
    \kx\vA^2\vxp ] \veck
   - \kx\vA^2(\veck\cdot\vecvp)\hatx =0\,,
\end{split}
\end{equation}
\citep[e.g.,][]{lit01}.  This is a generic dispersion relation for
waves in a uniform medium moving at a constant velocity.

\subsection{Canonical Modes}

As is well known, Equation (\ref{e:fdisp}) gives algebraic relations
for three different kinds of propagating waves: Alfv\'{e}n, fast, and
slow waves \citep[e.g.,][]{shu92}. In this subsection, we derive the
dispersion relation of each mode and the corresponding eigenvector with
arbitrary normalization, in a form suitable for our perturbation
analysis.

\subsubsection{Shear Alfv\'{e}n modes}\label{s:sAm}

Shear Alfv\'{e}n waves are transverse waves with $\veck \cdot \vecvp
=0$, for which Equation (\ref{e:fdisp}) is reduced to
\begin{equation}\label{e:disp_alf}
  \OD^2 + \kx^2\vA^2 =0\,.
\end{equation}
It then follows from Equations (\ref{e:con_p2})--(\ref{e:ind_p2})
that $\vecbp = \pm \vecvp$,
$(\veck\cdot\vecvp)=(\veck\cdot\vecbp)=0$, and $\vxp=\bxp=0$,
indicating that Alfv\'{e}n waves are incompressible and thus do not
rely on thermal pressure. The eigenvector of shear Alfv\'{e}n waves
is thus
\begin{equation}\label{e:Alfeig1}
  \Pp = 0\,,
\end{equation}
\begin{equation}\label{e:Alfeig2}
  \vecvp = (0, \kz, -\ky)\,,
\end{equation}
\begin{equation}\label{e:Alfeig3}
  \vecbp = (0, \pm \kz, \mp\ky)\,,
\end{equation}
where the signs distinguish two oppositely propagating waves. As we
will explain below, shear Alfv\'{e}n waves with $\kz\neq0$ are not
excited by the distortions of an IF that we impose and thus out of
consideration in our analysis.

\subsubsection{Fast and Slow modes}\label{s:fastslow}

When both the magnetic forces and the thermal pressure are
important, we take the scalar product of Equation (\ref{e:fdisp})
with $\veck$ and $\vecvA$. This results in two homogeneous linear
equations for $(\veck\cdot\vecvp)$ and $(\vecvA\cdot\vecvp)$. The
condition for the existence of non-trivial solutions yields
\begin{equation}\label{e:quar1}
  \OD^4 + k^2 (\cs^2 + \vA^2) \OD^2 + \kx^2k^2\cs^2 \vA^2 = 0\,,
\end{equation}
where $k=(\kx^2+\ky^2+\kz^2)^{1/2}$. This is a usual dispersion
relation for fast and slow MHD waves, for which $\vecvA$, $\vecvp$, and
$\veck$ lie in the same plane. We use Equations
(\ref{e:con_p2})--(\ref{e:div_p2}) to write the perturbed quantities in
terms of $\vzp$ as
\begin{equation}\label{e:eig1}
  \frac{\Pp}{\rho} = -\frac{(\OD^2 + k^2\vA^2)}{i\kz\OD}\vzp\,,
\end{equation}
\begin{equation}\label{e:eig2}
  \vecvp = \left(\frac{\kx (\OD^2 + k^2\vA^2)}{\kz\OD^2}, \frac{\ky}{\kz},
  1\right)\vzp\,,
\end{equation}
\begin{equation}\label{e:eig3}
  \vecbp = \left(\frac{-i(\ky^2+\kz^2)\vA}{\kz\OD},
    \frac{i\kx\ky\vA}{\kz\OD},
    \frac{i\kx\vA}{\OD}\right)\vzp\,,
\end{equation}
for $\OD\ne 0$ and $\kz \ne 0$. Note that the eigenvectors of fast and
slow modes with $\OD\ne 0$ can be completely specified by assigning one
variable (e.g., $\vzp$).

Equation (\ref{e:quar1}) has a special slow mode with $\OD=0$,
occurring when $\kx\vA=0$ (for $k\neq0$). Equations
(\ref{e:con_p2})--(\ref{e:div_p2}) then give
\begin{equation}\label{e:eig4}
  \ky \vyp + \kz \vzp = 0\,,
\end{equation}
and
\begin{equation}\label{e:eig5}
  \dfrac{\Pp}{\rho} + \vA\bxp = 0\,.
\end{equation}
This corresponds to {\it vortex} modes\footnote{The vortex mode is
  a special case of more general entropy-vortex modes
  \citep[e.g.,][]{lan59}.  In our formulation, the entropy perturbations
  are absent due to the choice of an isothermal equation of state.},
requiring no perturbation in the total pressure. These modes are
generated at a curved IF and then simply advected downstream from the
front. Although Equation \eqref{e:eig5} is a general expression for the
perturbed pressure of the vortex modes, the perturbations we consider
in the present work require that $\Pp=\bxp=0$ (see Section
\ref{s:mhd1}).

For later purposes, we introduce the following dimensionless
quantities:
\begin{equation}
 \sigma \equiv \frac{\Omega}{\vzu(\kx^2 + \ky^2)^{1/2}}\,,     \;\;\;
 \nu    \equiv \frac{i\kz}{(\kx^2 + \ky^2)^{1/2}}\,, \;\;\;
 \cos\psi   \equiv \frac{\kx}{(\kx^2 + \ky^2)^{1/2}} \,.
\end{equation}
Equation \eqref{e:quar1} can then be expressed as
\begin{equation}\label{e:quar2}
  \Mms^2\sigma_{D}^4 +
  \alphah^2(1-\nu^2)\sigma_{D}^2 +
  \alphah^4(1-\nu^2) \frac{\cos^2 \psi}{(1 + \beta^{-1})\MA^2} = 0\,,
\end{equation}
where $\sigma_D = \sigma + \alphah\nu$, and $\alphah \equiv \vz/\vzu =
1$ in the upstream side and $\alphah=\alpha$ in the downstream side.

\subsection{Perturbed Jump Conditions}

Perturbations given in Equation (\ref{e:fourier}) result from
sinusoidal distortions of an IF which would otherwise remain planar.
Since the IF involves discontinuities of fluid quantities, there are
certain conditions that the perturbation variables should obey at the
perturbed IF. Let the shape of a deformed IF be described by
\begin{equation}
  \mathcal{F}(x,y,z,t) =
  z - \zeta e^{\Omega t + i\kx x + i\ky y} = 0\,,
\end{equation}
where $\zeta \,(\ll k^{-1})$ denotes the amplitude of the IF
distortion. Then, the unit vector normal to the perturbed front is
given by $\hatn = \nabla \mathcal{F} /|\nabla \mathcal{F}| = \hatz -
i\kx\zeta\hatx - i\ky\zeta\haty$, while the two unit vectors
tangential to the front can be chosen as $\hat{\mathbf{t}}_{\rm a} =
\hatx + i\kx\zeta\hatz$ and $\hat{\mathbf{t}}_{\rm b} = \haty +
i\ky\zeta\hatz$, to the first order in $k\zeta$.

It is straightforward to show that in the frame comoving with the
perturbed IF, the velocity components parallel to $\hatn$,
$\hat{\mathbf{t}}_{\rm a}$, and $\hat{\mathbf{t}}_{\rm b}$ are given by
\begin{equation}\label{e:vperp1}
  v_{\rm n} = \vz + \vzp - \zeta\Omega\,,
\end{equation}
\begin{equation}
  \mathbf{v}_{\rm t}  =(\vxp + i\kx\zeta\vz, \vyp + i\ky\zeta\vz)\,,
\end{equation}
respectively.  Similar expressions for the magnetic fields read
\begin{equation}
  B_{\rm n} = \Bzp - i\kx\zeta\Bx\,,
\end{equation}
\begin{equation}\label{e:Bparallel2}
  \mathbf{B}_{\rm t}  = (\Bx + \Bxp, \Byp)\,,
\end{equation}
respectively.

Equations \eqref{e:con}--\eqref{e:ind} can be recast into
flux-conservative form and integrated over the small volume located on
the front. By applying the divergence theorem, we obtain the following
set of jump conditions
\begin{equation}\label{e:con_j}
  \Deltaf \left[ \rho v_{\rm n} \right] = 0\,,
\end{equation}
\begin{equation}\label{e:mom_j2}
  \Deltaf \left[ \rho v_{\rm n}\mathbf{v}_{\rm t} -
    \dfrac{B_{\rm n}\mathbf{B}_{\rm t}}{4\pi} \right] = 0\,,
\end{equation}
\begin{equation}\label{e:mom_j1}
  \Deltaf \left[ \rho v_{\rm n}^2 + P + \dfrac{B^2_{\rm t}}{8\pi} \right] =
  -\Deltaf \left[ \rho \right] g \zeta\,,
\end{equation}
\begin{equation}\label{e:ind_j}
  \Deltaf \left[ v_{\rm n}\mathbf{B}_{\rm t} - B_{\rm n}\mathbf{v}_{\rm t}
  \right] = 0\,,
\end{equation}
and
\begin{equation}\label{e:div_j}
  \Deltaf \left[ B_{\rm n} \right] = 0\,,
\end{equation}
(see, e.g., \citealt{shu92}). Here, $\Deltaf[f] \equiv f(z=\zeta +
0) - f(z=\zeta - 0 ) = f_2 - f_1$ indicates the difference of a
quantity $f$ evaluated at immediately behind and ahead of the front.

Substituting Equations \eqref{e:vperp1}--\eqref{e:Bparallel2} in
Equations \eqref{e:con_j}--\eqref{e:div_j}, one can show that the
zeroth-order terms lead to Equations
\eqref{e:con_st}--\eqref{e:ind_st}.  Taking the first-order terms, one
obtains
\begin{equation}\label{e:con_jp}
  \Deltaf \left[\rhop \vz + \rho (\vzp - \Omega\zeta) \right] = 0\,,
\end{equation}
\begin{equation}\label{e:momx_jp}
  \Deltaf \left[ \rho\vz(\vxp + i\kx\zeta\vz) - \rho\vA(\bzp - i\kx\zeta\vA)
  \right] = 0\,,
\end{equation}
\begin{equation}\label{e:momy_jp}
  \Deltaf \left[ \rho\vz(\vyp + i\ky\zeta\vz) \right] = 0\,,
\end{equation}
\begin{equation}\label{e:momz_jp}
  \Deltaf \left[ \rhop\vz^2 + 2\rho\vz(\vzp - \Omega\zeta) + \rho\vz\pp +
    \rho\vA\bxp \right] = -\Deltaf\left[ \rho \right] g\zeta\,,
\end{equation}
\begin{equation}\label{e:indx_jp}
  \Deltaf \left[ \rho^{1/2}(\bxp\vz + \vA(\vzp - \Omega\zeta)) \right] =
  0\,,
\end{equation}
\begin{equation}\label{e:indy_jp}
  \Deltaf \left[ \rho^{1/2}\byp\vz \right]=0\,,
\end{equation}
\begin{equation}\label{e:div_jp}
  \Deltaf \left[ \rho^{1/2}(\bzp - i\kx\zeta\vA) \right]=0\,,
\end{equation}
where $\pp = \Pp/(\rho\vz)$. Note that only two among Equations
\eqref{e:indx_jp}--\eqref{e:div_jp} are independent since the
induction equation automatically satisfies the divergence-free
condition for magnetic fields.\footnote{With help of the vertical
  component of Equation \eqref{e:ind_p2} and Equation
  \eqref{e:div_p2}, one can derive Equation \eqref{e:div_jp} directly
  from a linear combinations of Equations \eqref{e:indx_jp} and
  \eqref{e:indy_jp}.} Therefore, the perturbed jump conditions at the
front provide six constraints for the perturbation variables.

An additional constraint can be obtained by linearizing Equation
\eqref{e:pho} as
\begin{equation}\label{e:pho_p}
  \rhopu\vzu + \rhou(\vzpu - \Omega\zeta) = \mH \Fph^{\pr}\,,
\end{equation}
where $\Fph^{\pr}$ is the perturbed photon flux at the distorted IF.
For simplicity, we set $\Fph^{\pr} = 0$ in the present work, implying
that the mass flux per unit area through the IF is unchanged in the
perturbed state.  We note that $\Fph^{\pr}$  can be non-vanishing when
the effect of finite probability for photon absorption in the ionized
region is considered, suppressing the instability at scales larger than
the recombination length scale (see \citealt{axf64,wil02}).

Defining $Z \equiv \zeta (\kx^2 + \ky^2)^{1/2}\vzu$ and $\gb \equiv
g/((\kx^2 + \ky^2)^{1/2}\vzu^2)$, Equations
\eqref{e:con_jp}--\eqref{e:indy_jp} and \eqref{e:pho_p} can be
simplified as
\begin{equation}\label{e:pho_p2}
  \Msu^2\ppu + \vzpu - \sigma Z = 0\,,
\end{equation}
\begin{equation}\label{e:con_jp2}
  \Msd^2\ppd + \vzpd - \sigma Z = 0\,,
\end{equation}
\begin{equation}\label{e:momx_jp2}
  \vxpu - \MAu^{-1}\bzpu -\vxpd + \MAd^{-1}\bzpd + i(1 + \MAu^{-2} - \alpha(1 +
  \MAd^{-2}))\cos\psi Z = 0\,,
\end{equation}
\begin{equation}\label{e:momy_jp2}
\vypu - \vypd + i(1-\alpha)\sin\psi Z=0\,,
\end{equation}
\begin{equation}\label{e:momz_jp2}
  \left( (1 + \betau^{-1})(1-\Mmsu^2) +
    \cos^2\psi\dfrac{\MAu^{-2}(1- \alpha^{-2})}{(\sigma + \nu)^2}
  \right) \ppu - (1 + \betad^{-1})(1 - \Mmsd^2)\ppd + (1 - \alpha^{-1})\gb Z =
  0\,,
\end{equation}
\begin{equation}\label{e:indx_jp2}
  \bxpu - \MAu^{-1}\Msu^2\ppu - \alpha^{1/2}\left( \bxpd -
  \MAd^{-1}\Msd^2\ppd \right) = 0\,,
\end{equation}
\begin{equation}\label{e:indy_jp2}
  \bypu - \alpha^{1/2}\bypd = 0\,.
\end{equation}
Of the equations above, the derivation of Equation \eqref{e:momz_jp2}
is not trivial, requiring to utilize Equations
\eqref{e:quar1}--\eqref{e:eig3}: we present the necessary steps for it
in Appendix \ref{a:eqn}.

\section{Instability of Unmagnetized IFs}\label{s:hd}

We now want to explore the instability of an isolated, weak D-type IF.
Here, the term ``isolated'' implies that disturbances are generated
only at the front and decay at a large distance from the front. Of the
MHD waves described above, therefore, we consider only waves that are
evanescent away from the front, i.e., ${\rm Im}(\kz)< 0$ in the
upstream side ($z<0$) and ${\rm Im}(\kz) > 0$ in the downstream side
($z>0$), which is imposed by the regularity condition at infinity. In
our method, finding the growth rate as well as the eigenstate of
unstable modes takes two steps: (1) we express the perturbation
variables as a linear superposition of the canonical waves at each side
of the perturbed IF; (2) we then require the perturbation variables to
fulfill the jump conditions at the perturbed IF.

\citet{van62} was the first who studied the instability of
unmagnetized IFs. In this section, we revisit the problem to
exemplify our technique in the most simplest case, and to elucidate
the physical nature of the instability in analogy to the DLI. The
case of magnetized IFs will be presented in Section \ref{s:mhd}.

In the absence of magnetic fields, the background flows possess
rotational symmetry with respect to the $z$-axis, so that we may take
$\ky=0$ (hence $\cos\psi=1$) and $\vyp = 0$ without any loss of
generality. The solutions of Equation \eqref{e:quar2} in the limit of
$\MA$, $\beta \rightarrow \infty$ are
\begin{equation}\label{e:nua}
  \nua = \dfrac{1}{1-\Ms^2}\left[\Ms^2\dfrac{\sigma}{\alphah}
    \pm \sqrt{\Ms^2\left(\frac{\sigma^2}{\alphah^2} - 1\right)+1}\right]\,,
\end{equation}
\begin{equation}
  \nuv = -{\sigma}/{\alphah}\,,
\end{equation}
where the subscripts ``a'' and ``v'' stand for acoustic and vortex
modes, respectively, which are only modes that constitute the
perturbations at each side of the IF.  Since the vortex mode is
produced by the front deformation and then passively advected by the
background flows, it exists only in the downstream side.

Let $\mathbf{S} = (\pp,\,\vxp,\,\vzp)$ describe the eigenvectors of
the canonical waves such that
\begin{equation}\label{e:S_a}
  \mathbf{S}_{\rm{a}} = \left(- 1 - {\sigma}/{(\alphah\nua)}\,,
    {i}/{\nua},\, 1 \right),\,
\end{equation}
and
\begin{equation}\label{e:S_v2}
  \mathbf{S}_{\rm{v}} = \left( 0,\, -i\sigma/\alpha,\, 1 \right)\,.
\end{equation}
The boundary conditions for isolated IFs require that
$\rm{Re}(\nua)> 0$ in the upstream neutral region, while
$\rm{Re}(\nua) < 0$ in the downstream ionized region, as mentioned
above. For unstable modes with $\rm{Re}(\sigma) > 0$ and $\Ms^2 <
1$, one can write the total perturbations as a linear combination of
the canonical modes as
\begin{equation}\label{e:Sup}
  \mathbf{S}_{1} = C_{\rm a1}\mathbf{S}_{\rm{a1}}\,,
\end{equation}
in the upstream side, and
\begin{equation}\label{e:Sdown}
  \mathbf{S}_{2} = C_{\rm a2} \mathbf{S}_{\rm{a2}} + C_{\rm
  v2}\mathbf{S}_{\rm{v2}}\,,
\end{equation}
in the downstream side. Here, $C_{\rm a1}$, $C_{\rm a2}$, and
$C_{\rm v2}$ are the coefficients to be determined, and $\nua$ in
Equation \eqref{e:nua} should be calculated with the positive and
negative signs for $\mathbf{S}_{\rm{a1}}$ and
$\mathbf{S}_{\rm{a2}}$, respectively.

Plugging Equations \eqref{e:Sup} and \eqref{e:Sdown} into Equations
\eqref{e:pho_p2}--\eqref{e:momx_jp2} and \eqref{e:momz_jp2} for
$\vA=0$, we are left with a set of linear equations for four variables
$(C_{\rm
  a1},C_{\rm a2},C_{\rm v2}, Z)$. These can be cast into a matrix form
as
\begin{equation}\label{e:MatrixHD} {\small \left(\begin{array}{cccc}
        1 - \Msu^2\left(1 + \sigma/\nuau\right) & 0 & 0 & -\sigma \\
        0 & 1 - \Msd^2\left( 1 + \sigma/(\alpha\nuad) \right) & 1 & -\sigma \\
        \nuau^{-1} & -\nuad^{-1} & \sigma/\alpha & 1-\alpha \\
        -(1-\Msu^2)\left(1 + \sigma/\nuau \right) & (1
        -\Msd^2)\left(1 + \sigma/(\alpha\nuad) \right) & 0 & ( 1-
        \alpha^{-1})\gb
\end{array}\right)
\left(\begin{array}{c} C_{\rm a1} \\ C_{\rm a2} \\ C_{\rm v2} \\ Z
\end{array}\right)
 = 0\,.}
\end{equation}
In order to have a nontrivial solution, the $4\times4$ matrix in
Equation \eqref{e:MatrixHD} must have a vanishing determinant.  This
yields
\begin{equation}\label{e:disp1}
  (\sigma - \Gammad\alpha)\left[1 + \Gammau\left(\dfrac{\sigma}{\alpha} -
      \dfrac{\alpha-1}{\sigma}\right)\right] =
  (\alpha - \Gammad\sigma)\left[\sigma + \Gammau \left(1 + \dfrac{1-\alpha}{\alpha}
      \dfrac{\gb}{\sigma}\right)\right]\,,
\end{equation}
where
\begin{equation}
  \Gammau = [1 + \Msu^2(\sigma^2-1)]^{1/2}\,,
\end{equation}
\begin{equation}
  \Gammad = \left[1+\Msd^2(\sigma^2/\alpha^2-1)\right]^{1/2}\,,
\end{equation}
which is our desired dispersion relation for instability of
unmagnetized IFs.  Note that Equation \eqref{e:disp1} is the same as
Equation (79) of \citet[see also \citealt{byc08}]{van62} when the
direction of radiation is normal to the front.\footnote{The
conversion of symbols used in \citet{van62} to
  those in the present paper is $n \rightarrow
  \Msu\sigma$, $y_1 \rightarrow \Gammau$, and $y_2 \rightarrow
  \theta^{1/2}\Gammad$.
  }

\begin{figure}[!ht]
  \epsscale{1.}\plottwo{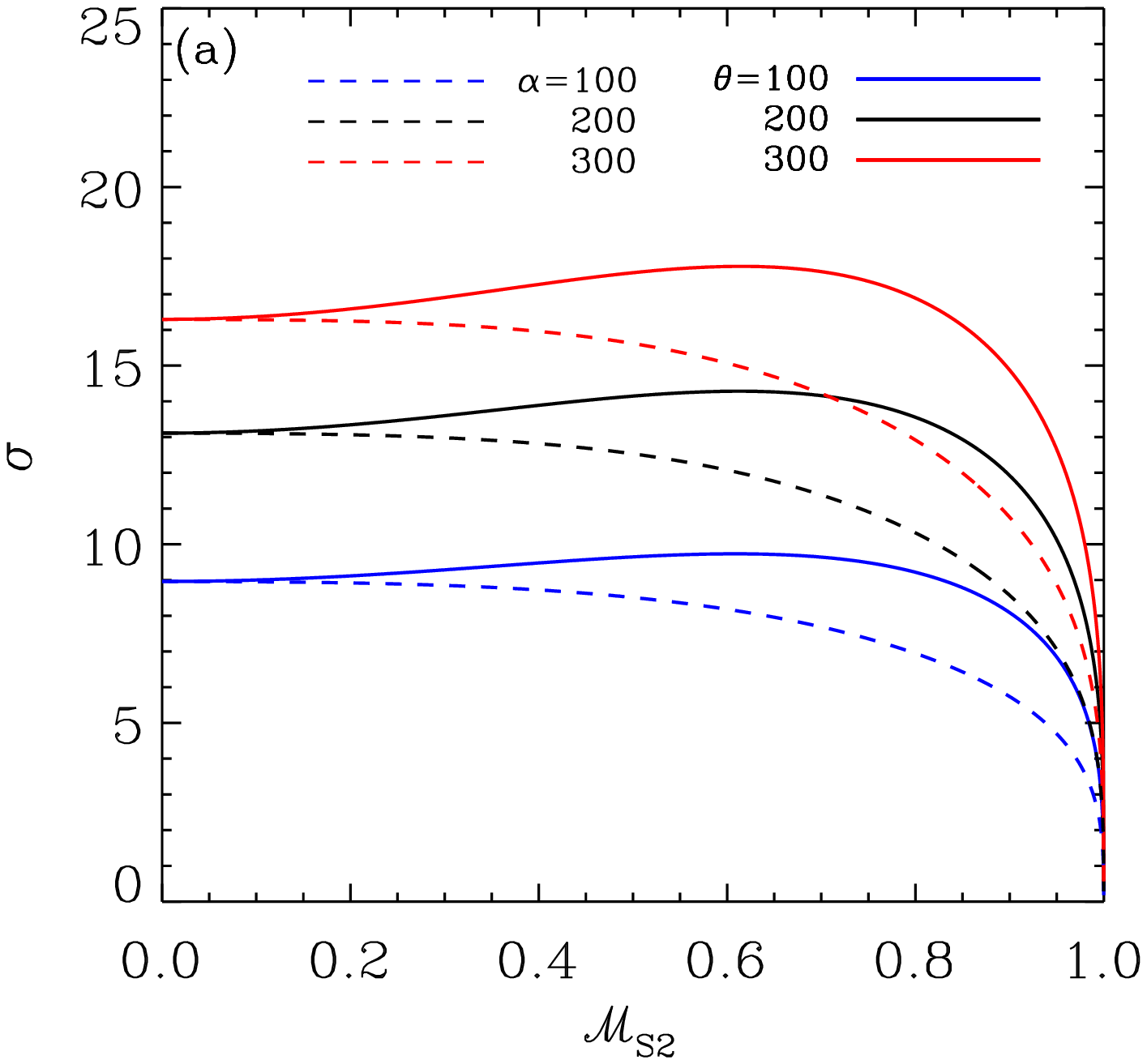}{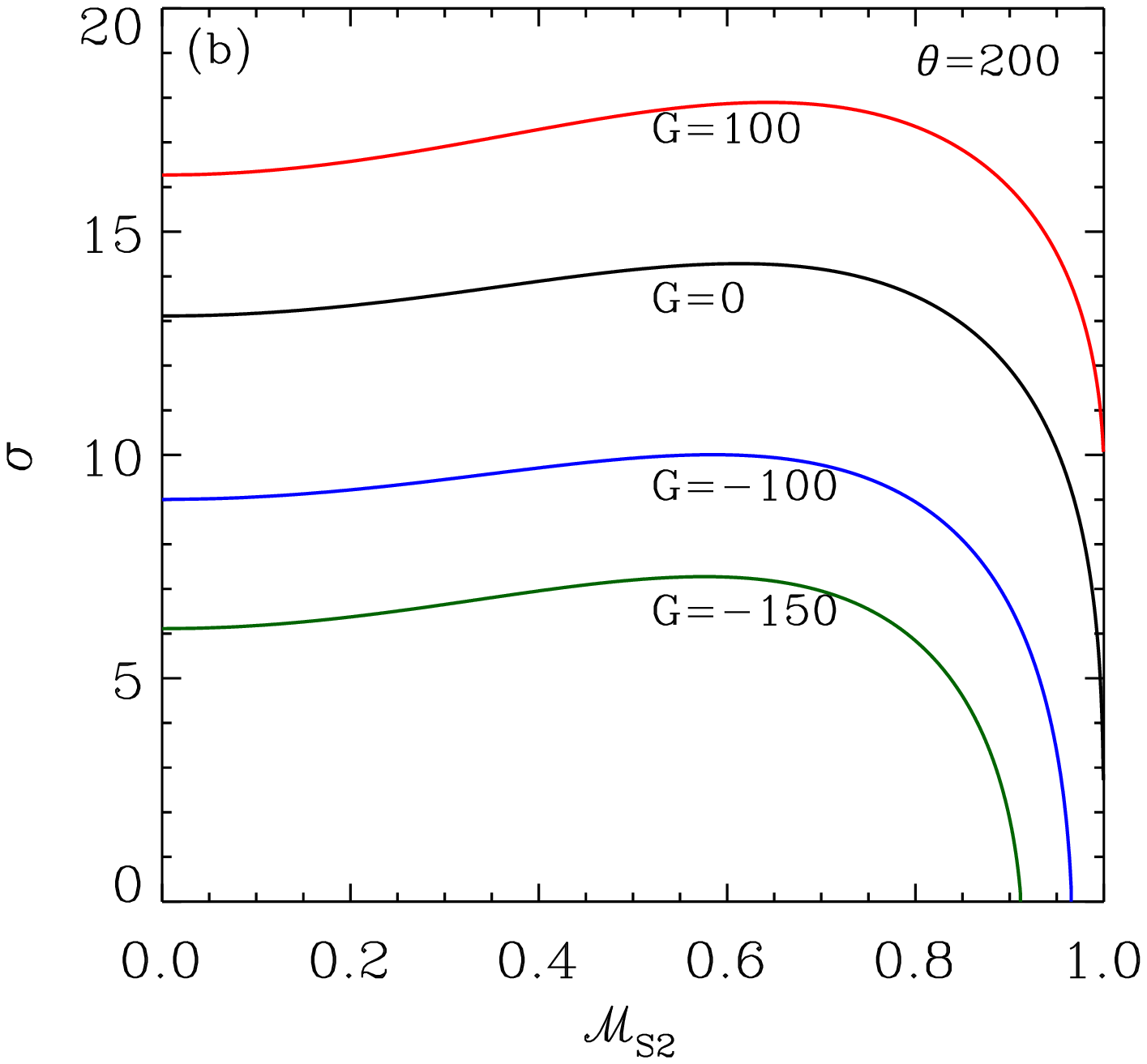}
  \caption{(a) Dimensionless growth rate $\sigma$ of the instability of
    unmagnetized IFs as a function of the downstream sonic Mach number $\Msd$
    for fixed heating factor $\theta$ (solid lines) and for fixed
    expansion factor $\alpha$ (dashed lines) with $\gb=0$. (b)
    Dependence of $\sigma$ on $G$ for $\theta=200$.  Accelerating
    fronts with larger $\gb$ are more unstable.}\label{f:03}
\end{figure}

In the incompressible limit of $\Ms \rightarrow 0$, Equation
\eqref{e:disp1} reduces to
\begin{equation}\label{e:disp2}
    \sigma = \dfrac{\alpha}{\alpha + 1}\left(\sqrt{1 + \alpha - \frac{1}{\alpha}
        +    \dfrac{\alpha^2 - 1}{\alpha^2}\gb} -1 \right)\,.
\end{equation}
In more general, compressible cases, however, Equation \eqref{e:disp1}
does not provide a closed-form expression for $\sigma$. Although it
can be converted to a polynomial by repeated squaring
\citep[e.g.,][]{sys97}, the resulting 16-th order polynomial in
$\sigma$ is not so illuminating that we present only the numerical
results here. Figure \ref{f:03}(a) plots as solid lines the
dimensionless growth rate $\sigma$ as a function of $\Msd$ for
$\theta=100$, $200$, and $300$, when $\gb=0$. For fixed $\theta$,
$\sigma$ increases slightly with increasing $\Msd \simlt 0.7$ due to
the increase in $\alpha$ (see Figure \ref{f:02}(b) and Equation
\eqref{e:disp2}). As $\Msd$ increases further, $\sigma$ starts to
decrease and tends to zero at $\Msd=1$ corresponding to the D-critical
front. Figure \ref{f:03}(a) also plots as dashed line $\sigma$ for
$\alpha=100$, 200, and 300, showing that $\sigma$ monotonically
decreases with increasing $\Msd$ for fixed $\alpha$. This suggests
that the stabilization of the IF instability is caused by gas
compressibility, as we will explain below.

Note that the incompressible dispersion of Equation \eqref{e:disp2}
with $\gb=0$ is identical to the dispersion relation of the DLI of an
evaporation front in an incompressible fluid
\citep[e.g.,][]{zel85,ino06,kim13}. Furthermore, Equation
\eqref{e:disp1} is equal to the full dispersion relation of the DLI
when the effect of compressibility is included \citep{byc08}. This
suggests that the physical nature of the instability of an IF is the
same as that of the DLI. When an IF is disturbed, the gas expansion
across the front makes the pressure drop (rise) on the part of the
distorted IF convex (concave) toward the ionizing source. The changes
in the pressure induce gas motions such that more (less) neutral gas is
directed toward to the convex (concave) parts. Since the ionizing
photon flux at the IF is assumed to be fixed, this makes the convex
(concave) parts advance further toward (recede away from) the ionizing
source in a runaway fashion, indicative of instability.

When gas compressibility is considered, the pressure drop (rise) is
partly translated into the drop (rise) in the perturbed density via
$\rhop/\rho \sim \Ms^{-2} \Pp/P$. This causes less changes in the
perturbed velocities compared to the incompressible
limit. Consequently, the amount of the perturbed mass flux at the
distorted IF is reduced, making the instability grow at a slower rate.
For the D-critical IF with $\Msd=1$, the perturbations in the
downstream side are unable to propagate into the upstream side since
the ionized gas is advected at the sound speed.  Accordingly, the
neutral gas does not respond to the deformation of the IF and remains
unperturbed. This can be seen more quantitatively from Equation
\eqref{e:momz_jp2} which gives $\ppu=0$ when $G=0$, which in turn
gives $C_{\rm a1}=0$ in Equation \eqref{e:Sup} hence $\sigma=0$ from
Equation \eqref{e:pho_p2}.

Figure \ref{f:03}(b) plots $\sigma$ for $\theta = 200$ and differing
$\gb$, showing the positive $\gb$ corresponding to an accelerating
front make the front more unstable. When the term involving $\gb$
dominates, Equation \eqref{e:disp2} recovers the growth rate of the
RTI. When $\gb> 0$, therefore, the DLI and RTI cooperate
constructively. For decelerating IFs with $\gb < 0$, on the other hand,
large scale modes with $k\vzu^2/|g| < \alpha$ are suppressed by
buoyancy.  The instability is completely quenched by buoyancy, provided
\begin{equation}\label{e:gbHD}
  \gb < - \alpha(1-\Msd^2)^{1/2}\,,
\end{equation}
which can be obtained by imposing $\sigma \rightarrow 0$ in Equation
\eqref{e:disp1}.\footnote{It has been known that an
accelerating ablation front is stabilized by thermal conduction (e.g.,
\citealt{byc94}). The relevant dispersion relation  is given by the
Takabe formula, $\Omega = a\sqrt{kg} - b kv_1$, where $k$ is
the perturbation wavenumber, $v_1$ is the velocity of the ablation
front, and $a \sim 0.9$ and $b \sim3$--$4$ are dimensionless
constants \citep{tak85}. The corresponding stability criterion written
in our notation reads $G = g/(kv_1^2) < b^2/a^2 \sim 10$.}

\section{Instability of Magnetized IFs}\label{s:mhd}

For the stability of magnetized IFs, we consider only two types of
perturbations: (1) perturbations with $\kx=0$ and $\ky \ne 0$ (i.e.,
$\cos\psi=0$) and (2) perturbations with $\kx \ne 0$ and $\ky = 0$
(i.e., $\cos \psi = 1$).  While these perturbations are not most
general, they can nevertheless capture the essential physics of
magnetic fields in the IF instability.

\subsection{Cases with $\kx=0$ and $\ky \ne 0$}\label{s:mhd1}

Perturbations with $\kx=0$ and $\ky \ne 0$ do not bend the field lines.
Since the front deformation does not involve the $x$-direction, we only
need to consider motions in the $y$-$z$ plane (i.e., $\vxp=0$).
Equations \eqref{e:ind_p2} then gives $\byp = \bzp = 0$, implying that
the perturbed magnetic fields exert magnetic pressure only in the
propagation direction of the disturbances and that shear Alfv\'en waves
are not excited.

For $\cos\psi = 0$, Equation \eqref{e:quar2} has two solutions
\begin{equation}\label{e:nuf}
  \nuf = \dfrac{1}{1-\Mms^2}\left[\Mms^2\dfrac{\sigma}{\alphah}
    \pm \sqrt{\Mms^2\left(\frac{\sigma^2}{\alphah^2} - 1\right)+1}\right]\,,
\end{equation}
\begin{equation}\label{e:nus}
  \nus = -{\sigma}/{\alphah}\,,
\end{equation}
where the subscripts ``f'' and  ``s'' stand for fast and slow modes,
respectively. Note that Equation \eqref{e:nuf} is identical to
Equation \eqref{e:nua} provided $\Ms$ is changed to $\Mms$. Note
also that Equation (\ref{e:nus}) is a dispersion relation for the
magnetized vortex modes with $\OD=0$, which exist only in the
downstream side from the IF, as explained in Section
\ref{s:fastslow}.

Now, let $\mathbf{S} = (\pp,\,\vyp,\,\vzp,\,\bxp)$ describe the
eigenvectors of the basis modes.  Then, Equations
\eqref{e:eig1}--\eqref{e:eig3} give
\begin{equation}\label{e:Sfu}
  \mathbf{S}_{\rm{f}} = \left[-\frac{1}{1 + \beta^{-1}}\left( 1 +
      \dfrac{\sigma}{\alphah\nuf}\right),\, \frac{i}{\nuf}, 1, -\frac{\Mms^2}{\MA}\left(1 +
      \dfrac{\sigma}{\alphah\nuf}\right) \right]\,,
\end{equation}
for the fast modes.  On the other hand, the eigenvectors of the slow modes in the
downstream side are given by
\begin{equation}
  \mathbf{S}_{\rm{s2}} = (0,\, -i\sigma/\alpha,\, 1,\, 0)\,,
\end{equation}
from Equation \eqref{e:eig4}. Note that here we take $\Pp=\bxp=0$ from
Equation \eqref{e:eig5} since $\Pp$ and $\bxp$ arising from the front
distortions should have the same sign when $\kx=0$.  Using the
condition that the waves should decay far away from the IF, one can
then write the state vectors as
\begin{equation}
  \mathbf{S}_{1} = C_{\rm f1}\mathbf{S}_{\rm{f1}}\,,
\end{equation}
\begin{equation}
  \mathbf{S}_{2} = C_{\rm f2} \mathbf{S}_{\rm{f2}} + C_{\rm s2}
  \mathbf{S}_{\rm{s2}}\,,
\end{equation}
where $C_{\rm f1}$, $C_{\rm f2}$, $C_{\rm s2}$ are coefficients to be
determined.

Following the same steps as in the case of unmagnetized IFs, we apply
the jump conditions (Equations \eqref{e:pho_p2}, \eqref{e:con_jp2},
\eqref{e:momy_jp2}, and \eqref{e:momz_jp2}) to obtain a linear system
of four equations in four unknowns $(C_{\rm f1}, C_{\rm f2}, C_{\rm
  s2}, Z)$.\footnote{Equation \eqref{e:indx_jp2} is automatically
  satisfied by our choice of the state vectors.} From the condition for
non-trivial solutions, we derive the dispersion relation for the
instability of the magnetized IFs with $\kx=0$ and $\ky \ne 0$, which
is identical to Equation \eqref{e:disp1}, provided $\Ms$ is replaced
by $\Mms$.

\begin{figure}
  \epsscale{0.9}\plotone{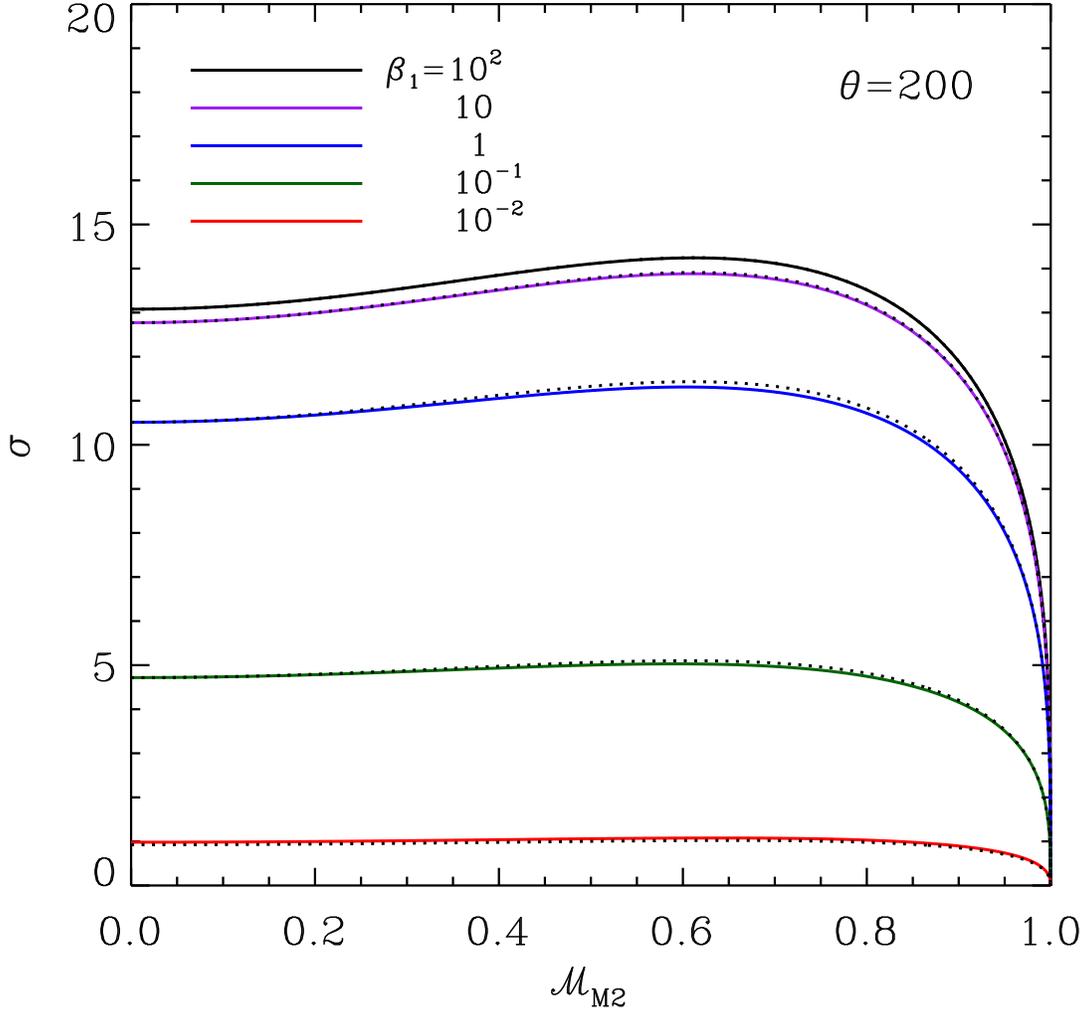}
  \caption{Growth rate of the instability of magnetized IFs with $\theta=200$ and
    differing $\betau$ as a function of $\Mmsd$, when the perturbations
    propagate perpendicular to the initial magnetic fields ($\kx = 0$).
    The solid lines plot the full numerical results of the magnetized cases,
    while the dotted lines draw the hydrodynamic counterparts (Equation \eqref{e:disp1})
    with $\alpha$ reduced by a factor of $1 + 1/(2\betau)$.}\label{f:04}
\end{figure}

Figure \ref{f:04} plots as solid lines the dimensionless growth rate
$\sigma$ for various $\betau$ when $\theta =200$ and $\gb=0$, as a
function of $\Mmsd$. While the overall shape of the dispersion
relations is unchanged compared to the hydrodynamics cases, magnetic
fields certainly reduce $\sigma$. The dotted lines plot the
unmagnetized dispersion relation, Equation \eqref{e:disp1}, with
$\alpha$ replaced by $\alpha/(1 + 1/(2\betau))$ which is the reduced
expansion factor due to magnetic fields (e.g., Equation
\eqref{e:mu_app}). The good agreement between the solid and dotted
lines suggests that the reduced growth rate in the magnetized case
results simply from a decreased $\alpha$ in the background state. When
$\kx=0$, magnetic fields remain straight and magnetized flows in the
linear regime behave similarly to unmagnetized flows, with fast
magnetosonic waves playing the exactly same role as acoustic waves.

\subsection{Cases with $\kx \ne 0$ and $\ky = 0$}\label{s:mhd2}

We examine the stability of magnetized IFs with respect to
perturbations lying in the $x$-$z$ plane, i.e., $\kx \ne 0$, $\ky = 0$,
and $\vyp = \byp =0$, for which not only magnetic pressure but also
magnetic tension affect the stability.  Note that these requirements
preclude the presence of shear Alfv\'en waves in the perturbations
(e.g., Equations \eqref{e:Alfeig2} and \eqref{e:Alfeig3}). We first
consider the incompressible limit and then generalize the results to
compressible cases.

\subsubsection{Incompressible Limit}\label{s:MHDin}

For simplicity let us take the limit $\cs \rightarrow \infty$ ($\rhop
= 0$), while the Alfv\'{e}n speed remains finite, so that $\beta
\rightarrow \infty$ and $\Mms \rightarrow 0$. Equation \eqref{e:quar2}
with $\cos \psi = 1$ then yields
\begin{equation}
  \nuppm = \pm 1\,,
\end{equation}
and
\begin{equation}\label{e:nuspm}
  \nuspm = -{\sigma}/{\alphah} \pm i\MA^{-1}\,,
\end{equation}
where the subscripts ``p'' and ``s'' refer to the potential modes and
slow (or pseudo-Alfv\'{e}n) modes, respectively. The potential modes
are a special case of the acoustic mode (or fast mode). For unstable
modes with Re$(\sigma)>0$, it is apparent that the wave motions in the
upstream side can be specified by the potential mode with $\nupp=1$,
while the other potential mode with $\nupm=-1$ and two slow modes with
$\nuspm$ (propagating in the opposite directions along the background
magnetic fields) coexist in the downstream side.

Defining $\mathbf{S} = (\pp,\, \vxp,\, \vzp,\, \bxp,\, \bzp)$, one can
use Equations \eqref{e:con_p2}--\eqref{e:div_p2} to construct the
eigenvector of each mode as
\begin{enumerate}
\item Upstream potential mode:
\begin{equation}\label{e:pot1}
  \mathbf{S}_{\rm p1} = \left( -(\sigma + 1),\, i,\, 1,\,
    -\dfrac{\MAu^{-1}}{\sigma + 1},\, \dfrac{i\MAu^{-1}}{\sigma +1}
    \right)\,,
\end{equation}
\item Downstream potential mode:
\begin{equation}\label{e:pot2}
  \mathbf{S}_{\rm p2} = \left( \dfrac{\sigma - \alpha}{\alpha},\, -i,\, 1,\,
    -\dfrac{\alpha\MAd^{-1}}{\sigma - \alpha},\, \dfrac{i\alpha\MAd^{-1}}{\sigma - \alpha}
    \right)\,,
\end{equation}
\item Downstream slow modes:
\begin{equation}\label{e:Alf}
  \mathbf{S}_{\rm s2\pm} = \left( \mp i\MAd\nuApm,\, i\nuApm,\,
    1,\, \pm i\nuApm,\, \pm 1 \right)\,,
\end{equation}
\end{enumerate}
where $\nuApm$ is the value of $\nuspm$ in Equation \eqref{e:nuspm}
evaluated at the downstream side. The total perturbations are then
given by
\begin{equation}
  \mathbf{S}_1 = C_{\rm p1} \mathbf{S}_{\rm p1}\,,
\end{equation}
and
\begin{equation}
  \mathbf{S}_2 = C_{\rm p2} \mathbf{S}_{\rm p2} +
                 C_{\rm s2+}\mathbf{S}_{\rm s2+} +
                 C_{\rm s2-}\mathbf{S}_{\rm s2-}\,,
\end{equation}
in the upstream and downstream sides, respectively.

Plugging these expressions into the perturbed jump conditions
(Equations \eqref{e:pho_p2}--\eqref{e:momx_jp2}, \eqref{e:momz_jp2},
and \eqref{e:indx_jp2}), one obtains a linear system of five equations
in five unknowns $(C_{\rm p1}, C_{\rm p2}, C_{\rm s2+}, C_{\rm s2-},
Z)$, which is given in a matrix form by
\begin{equation}
{\small  \left(\begin{array}{ccccc}
      1 & 0 & 0 & 0 & -\sigma \\
      0 & 1 & 1 & 1 & -\sigma \\
      1 - \dfrac{\MAu^{-2}}{\sigma + 1} &
      1 + \dfrac{\MAd^{-2}\alpha}{\sigma-\alpha} &
      -\nuAp - i \MAd^{-1} &
      -\nuAm + i \MAd^{-1} &
      1 + \MAu^{-2} - \alpha(1 + \MAd^{-2}) \\
      \mathcal{A} &
      -\dfrac{\sigma - \alpha}{\alpha} &
      i\MAd^{-1}\nuAp &
      -i\MAd^{-1}\nuAm &
      (1-\alpha^{-1})\gb \\
      -\dfrac{\MAu^{-1}}{\sigma + 1} &
      -\dfrac{\alpha^{3/2}\MAd^{-1}}{\sigma - \alpha} &
      -i\alpha^{1/2}\nuAp &
      i\alpha^{1/2}\nuAm &
      0
    \end{array}\right)
\left(\begin{array}{c}
C_{\rm p1} \\ C_{\rm p2} \\ C_{\rm s2+} \\ C_{\rm s2-} \\ Z
\end{array}\right)
= 0\,,}
\end{equation}
where $\mathcal{A} \equiv -(\sigma + 1)\left[ 1 + (1 - \alpha^{-2})
  \MAu^{-2}/(\sigma + 1)^2 \right]$.

We set the determinant of the matrix to zero to obtain the dispersion
relation
\begin{multline}\label{e:disp3}
  \sigma^3 + \dfrac{3\alpha + 1}{\alpha + 1}\sigma^2 - \dfrac{1}{\alpha(\alpha +
    1)}\left[\alpha^3 - \left(3 + \frac{1}{\MAu^2} - \gb\right)\alpha^2 - \gb\alpha -
  \frac{1}{\MAu^2}\right]\sigma\\ - \dfrac{\alpha - 1}{\alpha(\alpha + 1)}\left[\alpha^2 + \left(\gb -
  \frac{1}{\MAu^2}\right)\alpha + \MAu^{-2}\right] = 0\,.
\end{multline}
Note that Equation \eqref{e:disp3} is identical to Equation (112) of
\citet{dur04}, which is the dispersion relation for the incompressible
DLI in a magnetized gas.  This again demonstrates that the IF
instability and the DLI share the common physical origin. In the limit
of $\MAu \rightarrow \infty$, Equation \eqref{e:disp3} recovers
Equation \eqref{e:disp2} in the unmagnetized case. On the other hand,
we write Equation \eqref{e:disp3} in dimensional form and collect
lowest-order terms in $\vzu$ to obtain
\begin{equation}
  \Omega^2 = g\kx\dfrac{\alpha - 1}{\alpha + 1} - \kx^2\vAu^2
  \dfrac{1 + \alpha^2}{\alpha(\alpha + 1)}\,,
\end{equation}
which is the usual dispersion relation for the RTI of a magnetized
contact discontinuity, for which magnetic fields play a stabilizing
role \citep[e.g.,][]{cha61}.

\begin{figure}[!ht]
  \epsscale{.7}\plotone{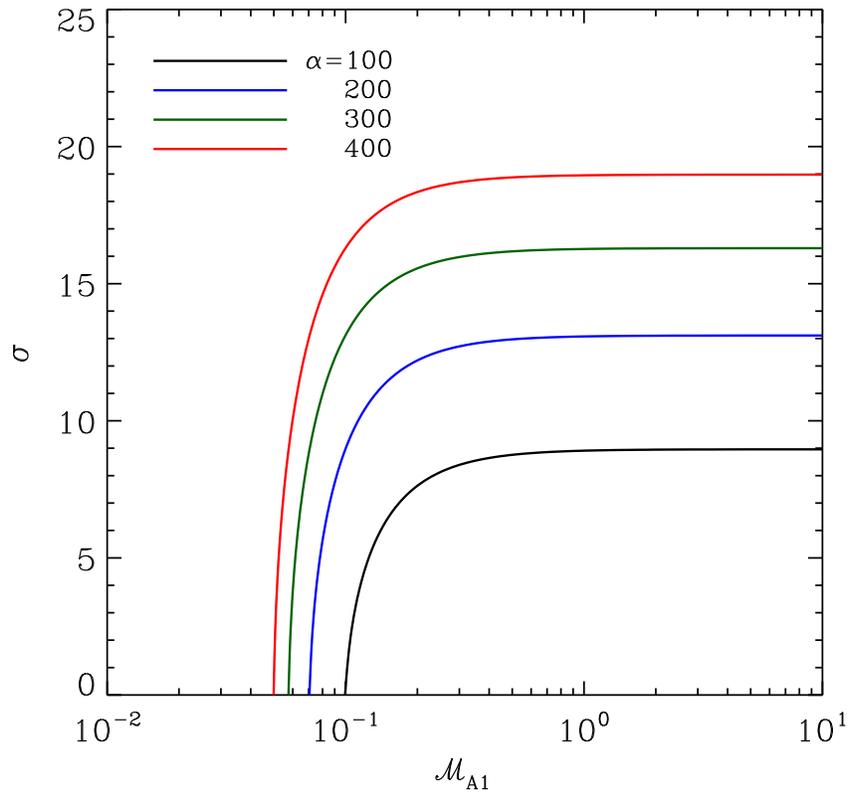}
  \caption{Incompressible growth rate $\sigma$ of the instability of magnetized IFs
    as a function of the upstream Alfv\'enic Mach number $\MAu$, for in-plane perturbations
    with $\ky=0$.
    For given $\alpha$, $\sigma$ becomes smaller as the field strength
    increases (or $\MAu$ decreases). The instability is
    completely suppressed at sufficiently small $\MAu$
    (see Equation \eqref{e:crit1}).}\label{f:05}
\end{figure}

For given $\alpha$ and $\MAu$, Equation \eqref{e:disp3} has only one,
if any, purely-growing solution with ${\rm Re}(\sigma)> 0$ and ${\rm
Im}(\sigma) = 0$, while the other two correspond to decaying solutions.
Figure \ref{f:05} plots the growth rate of the unstable mode for
various $\alpha$ as a function of $\MAu$, showing that $\sigma$
decreases as the field strength increases. It is a simple matter to
show that the instability becomes completely suppressed, provided
\begin{equation}\label{e:crit1}
  \MAu <   \sqrt{\dfrac{\alpha - 1}{\alpha(\alpha + \gb)}},\,\,\,\text{for
  stability}\,.
\end{equation}
The stabilization is due to magnetic tension forces that resist gas
motions across the field lines. This can be seen more quantitatively
as follows. The dimensionless growth rate with $\vA=\gb=0$ in the
incompressible limit is proportional to the fractional increase in the
mass flux, i.e., $\jzp/\jz = \vzpu/\vzu = \sigma \kx\zeta \approx
\sqrt{\alpha} \kx\zeta$ for $\alpha \gg 1$. The incoming velocity
change $\vzpu$ is due solely to the potential mode, which we denote by
$v_{z,\rm{p}}^{\pr}$. On the other hand, the distorted front deforms
the magnetic fields by an amount $\Bzp/\Bx \sim \kx\vzp/\Omega =
\kx\zeta$ from Equations \eqref{e:ind_p2} and \eqref{e:pho_p}. The
associated velocity change induced by magnetic tension is $|v_{z,\rm
  B}^{\pr}| = |\Bzp|/\sqrt{4\pi\rho} \sim \vA \kx\zeta$, which tends
to reduce $v_{z,\rm{p}}^{\pr}$ and hence $\jzp$. Note that
$|v_{z,\rm{p}}^{\pr}| \sim |v_{z,\rm B}^{\pr}|$ when $\MAu \sim
\alpha^{-1/2}$, entirely consistent with Equation \eqref{e:crit1} for
large $\alpha$.

\subsubsection{Compressible Cases}

We now consider more general compressible perturbations that are still
limited to the plane defined by the flow direction and magnetic fields
in the initial configuration. Although Equation \eqref{e:quar2}, a
quartic equation in $\nu$, has algebraic solutions, they are too
complicated for practical uses, so that we calculate the four
solutions numerically for given $\sigma$, $\alphah$, $\Mms$, and
$\MA$. From Equations \eqref{e:eig1}--\eqref{e:eig3}, the
corresponding eigenvector $\mathbf{S} = (\pp,\, \vxp,\, \vzp,\,
\bxp,\, \bzp)$ can be written as
\begin{equation}\label{e:fullS}
  \mathbf{S} = \left( \dfrac{-\chi}{\alphah\nu\sigma_{D}},
    \dfrac{i\chi}{\nu \sigma_{D}^2}, 1,
    \dfrac{-\alphah\nu\MA^{-1}}{\sigma_{D}},
    \dfrac{i\alphah\MA^{-1}}{\sigma_{D}} \right)\,,
\end{equation}
where $\chi \equiv \sigma_{D}^2 + \alphah^2\MA^{-2}(1-\nu^2)$.

Similarly to the incompressible case, two of the solutions represent
fast modes, while the remaining two are slow modes. For unstable modes
with $\rm{Re}(\sigma) > 0$, there is only one root with $\rm{Re}(\nu) >
0$ in the upstream side, which is a fast mode denoted by $\nu_{\rm
f1}$. On the other hand, the downstream side has three roots with
$\rm{Re}(\nu) < 0$: one pure real solution is a fast mode ($\nu_{\rm
f2}$)  and two complex roots are slow modes ($\nu_{\rm s2\pm}$). Upon
finding $\nu_{\rm f1}$, $\nu_{\rm f2}$, and $\nu_{\rm s2\pm}$, we
calculate the corresponding eigenvectors $\mathbf{S}_{\rm f1}$,
$\mathbf{S}_{\rm f2}$, and $\mathbf{S}_{\rm s2\pm}$ from Equation
\eqref{e:fullS}. We then construct the perturbations as
\begin{equation}\label{e:s1_MHDc}
  \mathbf{S}_1 = C_{\rm f1}\mathbf{S}_{\rm f1}\,,
\end{equation}
and
\begin{equation}\label{e:s2_MHDc}
  \mathbf{S}_2 = C_{\rm f2}\mathbf{S}_{\rm f2} + C_{\rm s2+}\mathbf{S}_{\rm
    s2+} + C_{\rm s2-}\mathbf{S}_{\rm s2-}\,,
\end{equation}
in the upstream and downstream sides, respectively, with the unknown
coefficients $C_{\rm f1}$, $C_{\rm f2}$, and $C_{\rm s2\pm}$.

Substituting Equations \eqref{e:s1_MHDc} and \eqref{e:s2_MHDc} in
Equations \eqref{e:pho_p2}--\eqref{e:momx_jp2} and
\eqref{e:momz_jp2}--\eqref{e:indx_jp2}, we obtain a set of five linear
equations in five unknowns ($C_{\rm f1}$, $C_{\rm f2}$, $C_{\rm s2+}$,
$C_{\rm s2-}$, $Z$).  The resulting equation in a matrix form is
displayed in Appendix \ref{a:mat}. To obtain non-trivial solutions, we
set the determinant of the matrix $\mathbf{A}$ in Equation
\eqref{e:mat} equal to zero.  To calculate $\sigma$ numerically, we
first take trial values for the real and imaginary parts of $\sigma$
and calculate four $\nu$'s, ensuring that the perturbed flow decays
away from the front. We then check if the determinant vanishes or not.
If the determinant is not sufficiently small, we return to the first
step and change $\sigma$. We repeat the iterations until the converged
solutions are obtained within tolerance of $10^{-6}$. We have confirmed
that our numerical method gives the same dispersion relations as
Equations \eqref{e:disp1} for unmagnetized cases and \eqref{e:disp3}
for $\cs^2 \gg \vz^2, \vA^2$.

\begin{figure}[!ht]
  \epsscale{1.}\plottwo{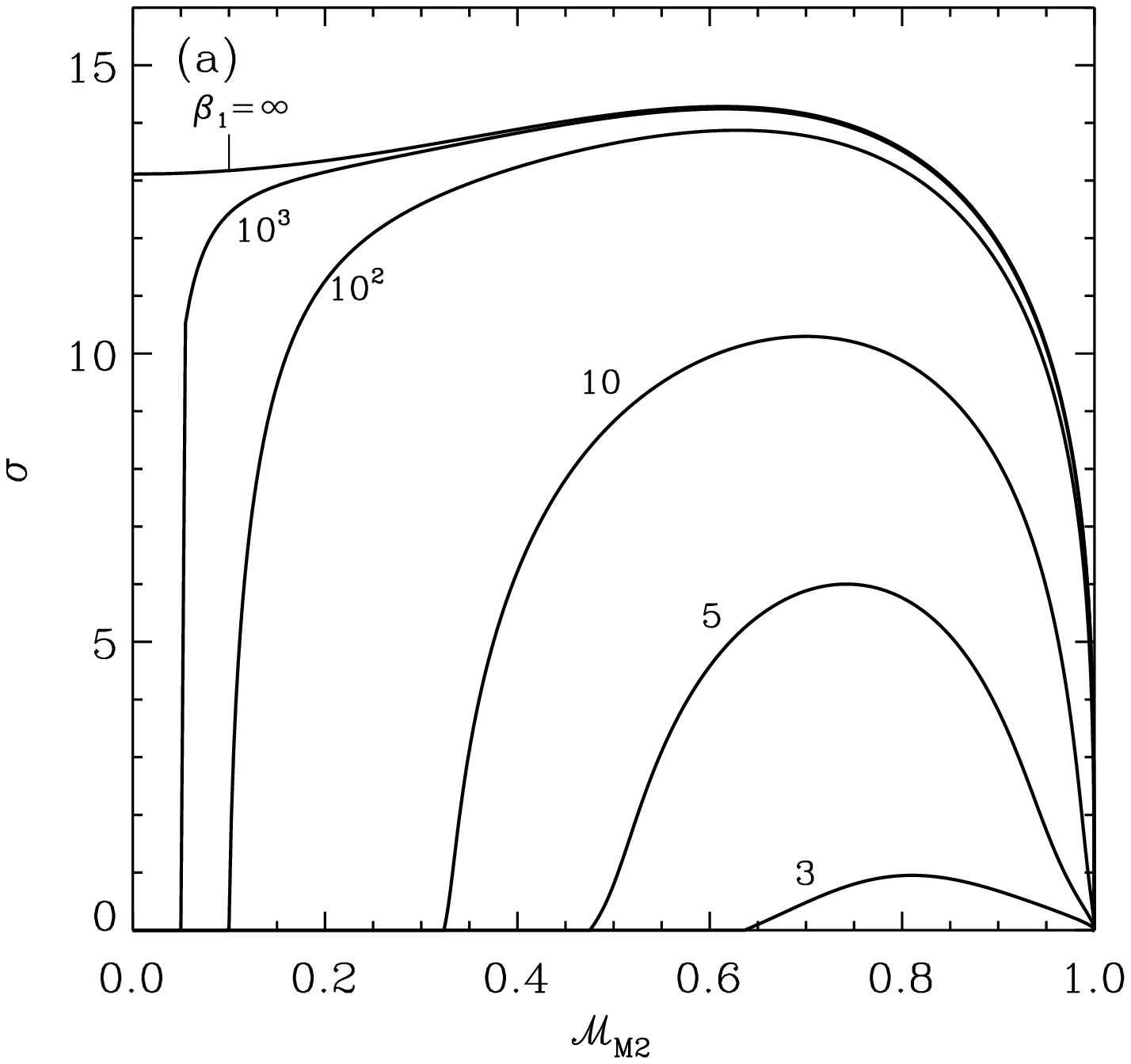}{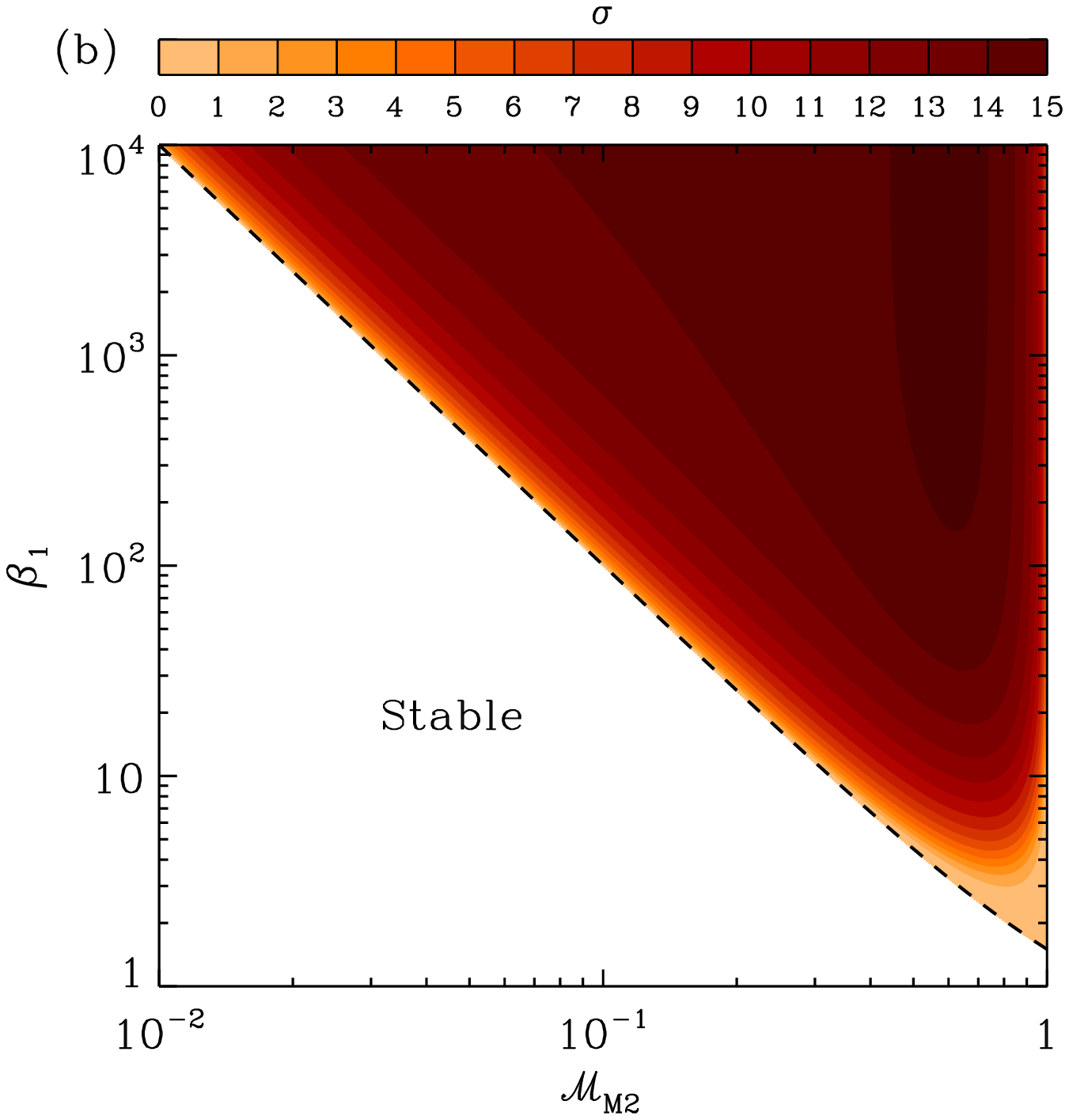}
  \caption{(a) Compressible growth rate $\sigma$ of the instability of
    magnetized IFs as a function of $\Mmsd$ and (b) the contour of
    $\sigma$ in the $\betau$-$\Mmsd$ plane. The perturbations are limited
    to $\ky=0$, and $\theta=200$ and
    $\gb=0$ are chosen. The instability is
    suppressed by magnetic tension at low $\Mmsd$ and by
    compressibility at $\Mmsd \rightarrow 1$.
    The black dashed line in (b) draws the stability criteria (Equation \eqref{e:crit2})
    in the limit $\theta \gg 1$, in good agreement with the full
    numerical results.
    }\label{f:06}
\end{figure}

We find that $\sigma$ of the unstable modes is pure real, as in the
incompressible case, and goes to zero for the D-critical IF regardless
of $\betau$. Figure \ref{f:06}(a) plots the resulting growth rates as a
function of $\Mmsd$ for $\theta = 200$ and $\gb=0$ but differing
$\betau$, while Figure \ref{f:06}(b) plots contours of $\sigma$ in the
$\betau$-$\Mmsd$ plane.  For $\betau \gg 1$, the growth rates are not
much different from the unmagnetized counterparts, except for $\Mmsd
\ll 1$. As the magnetic field strength increases, however, not only do
the growth rates decrease but also the unstable range of $\Mmsd$
shrinks. For small $\Mmsd$, the magnetic tension forces stabilize the
instability, as in the incompressible limit. For $\alpha\gg1$ and
$G=0$, Equation \eqref{e:crit1} with the help of Equations
\eqref{e:mu_app} and \eqref{e:Mmsu_app} can be written as
\begin{equation}\label{e:crit2}
  \Mmsd < \left(\dfrac{2}{2\betau - 1}\right)^{1/2},\,\,\,\text{for
  stability},
\end{equation}
indicating that the instability of magnetized IFs is completely
suppressed by magnetic tension when $\betau \leq 3/2$. The dashed line
in Figure \ref{f:06}(b) draws Equation \eqref{e:crit2}, in excellent
agreement with the stability criterion found numerically for the whole
range of $\Mmsd$.

\section{Summary and Discussion}\label{s:sum}

\subsection{Summary}

We have performed a linear stability analysis of magnetized, weak
D-type IFs around \ion{H}{2} regions.  This work extends \citet{van62}
who analyzed the stability of IFs in the absence of magnetic
fields. To simplify the situation, we consider an IF in plane-parallel
geometry, perpendicular to the incident direction of ionizing photons,
and ignore the effects of recombination in the ionized gas in the
present work. We further assume that magnetic fields are parallel to
the front and that the gas remains isothermal with different
temperatures in the neutral and ionized sides of the front. We first
solve for equilibrium configurations of steady IFs across which total
mass, momentum, and magnetic fluxes are conserved. We find that for
weak D-type IFs, magnetic fields tend to increase the maximum
propagation speed of the IFs, while reducing the expansion factor
$\alpha$ by a factor of $1+1/(2\betau)$ compared to the unmagnetized
case (see Equations \eqref{e:mu_app} and \eqref{e:Mmsu_app}). In the
stationary IF frame, the magnetosonic Mach number of the ionized gas
downstream from the IF always satisfies $\Mmsd^2\leq 1$ for weak
D-type fronts, with the equality corresponding to the D-critical
fronts. We provide the approximate expressions (Equations
\eqref{e:crM2} and \eqref{e:crA2}) for the expansion factors and sonic
Mach numbers when the fronts are either D- or R-critical.

We impose small-amplitude perturbations on a steady-state IF in
isolation, and seek for unstable modes that grow exponentially in time.
The perturbations are constructed as a superposition of MHD waves that
are evanescent far away from the IF; only the fast (or acoustic) mode
propagates in the upstream side, while both fast and slow (or vortex)
modes exist in the downstream side for an isolated magnetized (or
unmagnetized) front.  For the two-dimensional perturbations we impose,
shear Alfv\'en waves are not excited. We require that the perturbation
variables satisfy the perturbed jump conditions (Equations
\eqref{e:pho_p2}--\eqref{e:indy_jp2}) at the IF to derive the
dispersion relations of the instability.

We first apply our technique to unmagnetized IFs.  The resulting
dispersion relation (Equation \eqref{e:disp1}) recovers the result of
\citet{van62}. When the external gravity is ignored (i.e., $G=0$), it
is also identical to the dispersion relation of the DLI seen in laser
ablation fronts in inertial confinement fusion
\citep[e.g.,][]{byc08,mod09}. This suggests that the physical
mechanism behind the IF instability is the same as that of the
DLI. The DLI is generic for any interfacial layer through which a cold
dense gas absorbs heat and expands to turn to a warm rarefied gas. The
DLI is multi-dimensional, requiring wavy deformation of the interface
in the direction normal to the incident ionizing radiation. In the
case of IFs, the front deformation grows due to mismatches between the
perturbed mass flux and the ionization rate of the cold gas at the IF.
This is in contrast to the claim of \citet{van62} that the IF
instability is due to the rocket effect \citep{kah54, oor55} which,
unlike the DLI, does not require wavy deformation of the IF, and
relies on high-speed evaporating gas from a dense cloud to exert
thrust on it.

The unstable mode of the IF instability grows without oscillation,
indicative of pure instability. The growth rate scales linearly with
the wavenumber as well as the background fluid velocity relative to the
front. The dimensionless growth rate increases with $\alpha^{1/2}$ for
$\alpha \gg 1$ (Equation \eqref{e:disp2}). As $\Msd$ increases, the
instability is stabilized by gas compressibility which tends to reduce
the change in mass flux at the IF, becoming completely quenched when
the front is D-critical ($\Msd=1$). The IF instability cooperates with
the RTI for IFs accelerating away from an ionizing source ($G>0$),
while it is suppressed by buoyancy for decelerating IFs at large scales
(Equation \eqref{e:gbHD}).

For magnetized fronts, we consider two cases of two-dimensional
perturbations: (1) perturbations with $\kx=0$ are in the plane
perpendicular to the magnetic fields and (2) perturbations with $\ky=0$
are confined to the plane defined by the magnetic fields and the
background flows.  For the $\kx=0$ perturbations, the perturbed fields
exert only magnetic pressure forces and the resulting dispersion
relation is identical to the hydrodynamic case, provided the sound
speed is replaced by the speed of magnetosonic waves. For the $\ky=0$
perturbations, on the other hand, magnetic tension from the bent field
lines stabilizes the instability.  In the incompressible limit, the
dispersion relation (Equation \eqref{e:disp3}) of the IF instability is
again the same as that of the DLI studied by \citet{dur04}. The IF
instability is completely suppressed if the Alfv\'{e}nic Mach number is
sufficiently small (Equation \eqref{e:crit2}), suggesting that no
instability is expected if the plasma parameter $\betau$ is less than
3/2 in the upstream neutral region.

\subsection{Discussion}

Observations indicate that IFs are usually magnetized.  Using the
Zeeman effects of \ion{H}{1} and OH lines, for instance, \citet{bro99}
and \citet{bro01} reported that the strength of line-of-sight magnetic
fields toward the interface of \ion{H}{2} region/molecular cloud
complex in M17 is on average $\sim 200$--$500 \,\mu$G and reaches a
value as high as $\sim 750\,\mu$G. Taking $B_1 \sim 300\,\mu{\rm G}$,
corresponding to the magnetic pressure support of $2.6 \times 10^7 \kB
\,{\rm cm^{-3}\, K}$, and assuming $T_1 = 100 \Kel$, $\betau$ is less
than 1.5 unless $n_1 \simgt 8 \times 10^5 \;\rm{cm}^{-3}$. This large
background density is highly unlikely since the equilibrium model of
\citet{pel07} favors $n_1$ of order of $\sim 10^4\, {\rm
  cm^{-3}}$. This suggests that the IF instability is readily
stabilized by magnetic tension in the direction parallel to the fields.
However, the IFs in M17 can still be unstable to perturbations (with
$\kx=0$ and $\ky\neq0$) propagating in the direction perpendicular to
the fields. The growth rate $\Omega = \ky\vzu\sigma$ in dimensional
units can be written as
\begin{equation}
  \Omega \approx 6.4\times 10^{-5} \frac{\Mmsu\sigma}
  {(1 + \betau^{-1})^{1/2}}
  \left(\dfrac{\lambda}{0.1\;{\rm pc}}\right)
  \left(\dfrac{\csu}{1\;{\rm km\,s}^{-1}}\right) \;{\rm yr^{-1}}\,,
\end{equation}
where $\lambda$ is the perturbation wavelength.  For $G=0$ and $\theta
= 200$, the maximum value of $\Mmsu\sigma$ is $(0.19, 0.3, 0.40)$ for
$\betau=(0.01, 0.1, 1)$, which occurs at $\Mmsd = 0.75$ (see Figures
\ref{f:02}(a) and \ref{f:04}). The corresponding $e$-folding growth
time of the instability is $(2.8, 1.5, 0.8) \times 10^4\,{\rm yr}$ for
$\betau=(0.01, 0.1, 1)$, respectively, which is an order of magnitude
shorter than the typical expansion time scale of \ion{H}{2} regions
(typically $10^5\text{--}10^6\,{\rm yr}$; see below). This indicates
that the IF instability with $\ky\neq0$ can grow significantly, but by
keeping the magnetic field lines straight.

The relative importance of the front acceleration/deceleration to the
the gas kinetic energy is measured by the dimensionless parameter $\gb
= g/(k\vzu^2)$. Recently, \citet{ric14} investigated the stabilizing
effect of recombination on the RTI of accelerating IFs assuming that
gas is incompressible. Due to the incompressibility assumption,
however, his results (Equations (40) and (41) of \citep{ric14}) in the
absence of recombination recovers the dispersion relation only for the
RTI, but is unable to capture the IF instability of \citet{van62}. A
simple comparison between the growth rates of the IF instability ($\sim
\sqrt{\alpha}\vzu\kx$) and the RTI ($\sim \sqrt{g\kx}$) suggests that
the perturbed flows are strongly affected by buoyancy for $|G| \simgt
\alpha$. In a uniform medium without magnetic fields, \citet{spi78}
showed that a D-type IF expands as $r_{\rm IF} \approx r_{s} (1 +
7t/4t_s)^{4/7}$, where $r_{s} = 3.2 \,{\rm pc} (Q/10^{49}\,{\rm
s^{-1}})(\rhod/(100\,\mH\cm^{-3}))^{-2/3}$ is the initial Str\"{o}mgren
radius, $Q$ is the ionizing photon luminosity of a central star, and
$t_s = r_{s}/\csd = 0.24\,{\rm Myr}$ for $\csd = 13\,{\rm km\,s^{-1}}$.
The effective gravity is then $g = -3\csd/(4t_s) (1 +
4t/7t_s)^{-10/7}$.  Taking $\theta = 200$, $g = -3\csd/4t_s$, and
$\Mmsd = 0.5$, the IF instability is stabilized by buoyancy for
perturbations with wavelength $\lambda/r_s> 8\pi \Msd^2 /(3\alpha) \sim
10^{-2}$ even for purely hydrodynamic IFs. On the other hand, for IFs
accelerating outward in a stratified medium with density decreasing
more steeply than $r^{-3/2}$ \citep[e.g.,][]{fra90,wha08}, the RTI
would work together with the IF instability to make the fronts more
unstable.

We have not considered thermal conduction in our analysis since its
effect on the IF instability is thought to be insignificant. This can
be seen quantitatively as follows.  Let us take typical values for the
mean photoionization cross section $\sigpi = 3\times 10^{-18} \cm^2$,
the mean kinetic energy of a photo-ejected electron $\langle E_{\rm
  ph}\rangle = 2.4 \eV$ \citep[e.g.,][]{wha04}, and the incident
ionizing photon flux $F_{\rm ph} = \rhod\vzd/\mH$.  Then, the
volumetric heating rate by photoionization amounts to $\mathcal{H} =
(\rhod/\mH)\sigpi F_{\rm ph}\langle E_{\rm ph}\rangle = 1.5\times
10^{-19} (\rhod/(100\,\mH \cm^{-3})) (\csd/13\,{\rm km\,s^{-1}}) \Msd
\;{\rm erg}\,{\rm cm}^{-3}\,{\rm s}^{-1}$.  For the Spitzer
conductivity of $\kappa_{\rm Sp} \sim 10^4 \condunit$ in the ionized
gas with $T_2=10^4 \Kel$ \citep{mck77}, the Field length is calculated
to be $L_{\rm{F2}} = \sqrt{\kappa_{\rm Sp}T_2/\mathcal{H}} = 8.4
\times 10^{-6} \pc$, while the thermal diffusion length is
$L_{\rm{D2}} = \kappa_2/(\jz c_P) = 2 \times 10^{-5} \pc$, where $c_P$
is the specific heat at constant pressure (e.g.,
\citealt{kim13}). Note that the photon mean-free path $L_{\rm{mfp2}} =
1/(n_2\sigpi) = 10^{-3} \pc$ is much larger than the conduction length
scales. Therefore, thermal diffusion is unlikely to be important in
determining the structures of IFs as well as the IF instability (see
also \citealt{spi78}). This is in contrast to the case of evaporation
fronts between cold and warm gases studied by \citet{kim13}, where
thermal conduction not only affects the front thickness but also
stabilizes the DLI at small scales.

Can the IF instability manifest in numerical simulations of expanding
\ion{H}{2} regions?  In Eulerian simulations, it is unavoidable to
have spurious numerical viscosity, caused by a finite difference
scheme, that dampens perturbations at small scale. \citet{kim08} found
that the numerical diffusivity can be written as $\eta = A_d v_{\rm
  ad}\Delta x(\Delta x/\lambda)^n$, where $A_d$ is a dimensionless
constant, $\lambda$ is the characteristic length scale of
perturbations, $\Delta x$ is the grid spacing, $v_{\rm ad}$ is the
advection velocity through a numerical grid, and $n$ is the order of
the spatial reconstruction in the numerical scheme. Since the
corresponding damping time over the scale of $\lambda$ is $\tau_{\eta}
\sim \lambda^2/\eta$, the numerical diffusion would prohibit the
growth of the IF instability if $\tau_\eta \Omega \lesssim 1$ or if
$\lambda/\Delta x \lesssim (A_d\alpha v_{\rm ad} / 2\pi
\vzd\sigma)^{1/(n+1)}$. Taking $A_d = 8.1\times 10^4$ and $n=2$ from
\citet{kim12} for the {\it Athena} code with piecewise-linear
reconstruction scheme, for example, and taking $\alpha/\sigma \sim 10$
and $v_{\rm ad}/\vzd =1$, one can see that perturbations with
$\lambda/\Delta x \gtrsim 50$ would be stabilized by numerical
effects. Considering the typical resolution of $\Delta x \sim
10^{-2}\text{--}10^{-1} \,\pc$ in simulations of a single \ion{H}{2}
region \citep[e.g.,][]{kru07,mac11,art11}, the DLI mode below $\sim
0.1$--$1\,\rm{pc}$ would be suppressed by numerical diffusion. Modes
with wavelength longer than these may still grow, but too slowly to be
readily evident in numerical simulations. Therefore, very
high-resolution simulations are desirable to resolve the IF
instability at small scales.

Finally, we remark a few caveats made by our simplified model of an
IF.  First, the approximation of stationary IFs in plane-parallel
geometry ignores the curvature effect as well as temporal changes in
the background state, while IFs in reality have non-vanishing
curvature, especially in the case of ablated globules. As an IF
propagates, its curvature and the Mach number of the inflowing neutral
gas would vary. For an expanding \ion{H}{2} region, the perturbation
wavelength increases in proportion to the size of the \ion{H}{2}
region, resulting in a power-law growth rather than an exponential
growth \citep{zel85}. Second, IFs associated with blister-type
\ion{H}{2} regions and cometary globules (e.g.,
\citealt{kah69,ber89,ber90}) are strong D-type (or D-critical) and the
ionized gas accelerates away from such IFs to achieve a supersonic
speed. With spatially-varying density and velocity fields in the
background state, instability of strong D-type IFs cannot be explored
by our current technique that assumes uniform backgrounds in the
upstream and downstream sides. Third, a D-type IF is usually preceded
by a shock front, indicating that some waves launched by the IF would
undergo reflection at the shock front, which is likely to alter the
modal behavior of the perturbations in the upstream side. It is well
known that after a shock breakout, the shocked layer is subject to a
thin-shell instability due to the force imbalance between thermal and
ram pressures at the boundaries \citep{giu79,vis83,gar96}. Right after
the shock breakout (i.e., when the fronts are near D-critical), the
gas ahead of the IF is subject to significant non-steady cooling and
heating, so that the isothermal approximation in the neutral gas may
not be applicable \citep{hen09}. In this case, a proper account of
radiative cooling/heating is necessary to assess the impact of
nonlinear development of the IF instability \citep{wha08}.

\acknowledgments We thank the anonymous referee for careful reading of
our manuscript and many insightful comments and suggestions.  This
work was supported by the National Research Foundation of Korea (NRF)
grant, No. 2008-0060544, funded by the Korea government (MSIP).  The
work of J.-G.K. was supported by the National Research Foundation of
Korea (NRF) grant funded by the Korean Government (NRF-2014-Fostering
Core Leaders of the Future Basic Science Program).

\appendix

\section{Derivation of Equation \eqref{e:momz_jp2}}\label{a:eqn}

Here we present steps to derive Equation \eqref{e:momz_jp2}. Using
Equations \eqref{e:pho_p2} and \eqref{e:con_jp2}, Equation
\eqref{e:momz_jp} is reduced to
\begin{equation}\label{e:App1}
  (1 - \Msu^2)\ppu + \MAu^{-1}\bxpu - \MAd^{-1}\bxpd + (1 -
  \alpha^{-1})\gb Z
  = (1 - \Msu^2)\ppd\,.
\end{equation}
Combining Equations \eqref{e:indx_jp2} and \eqref{e:App1}, one may
write
\begin{equation}\label{e:App2}
  (1 + \betau^{-1} - \Msu^2)\ppu + (1-\alpha^{-2})(\MAu^{-1}\bxpu -
  \betau^{-1}\ppu) + (1 - \alpha^{-1})\gb Z = (1 + \betad^{-1} -
  \Msd^2)\ppd\,.
\end{equation}
As we shall show in Sections \ref{s:hd} and \ref{s:mhd}, the
perturbations in the upstream side (region 1) are described only by
fast modes that obey Equations \eqref{e:eig1}--\eqref{e:eig3}. Hence,
one can write for fast modes
\begin{align}
    \MAu^{-1}\bxpu - \betau^{-1}\ppu & = \left(
    \dfrac{\MAu^{-2}(\nu^2 - \sin^2\psi)}{\chi} -
      \frac{1}{\betau} \right)\ppu\,, \\
      & = \cos^2\psi\dfrac{\MAu^{-2}}{(\sigma + \nu)^2}\ppu\,, \label{e:App3}
\end{align}
where $\chi = \sigma_D^2 + \alphah^2\MA^{-2}(1 - \nu^2)$ and we have
used Equation \eqref{e:quar2} in the last equality.  Combining
Equations \eqref{e:App2} and \eqref{e:App3} gives Equation
\eqref{e:momz_jp2}.

\section{Perturbation Equations for $\kx\neq0$ and $\ky=0$}\label{a:mat}

We plug Equations \eqref{e:s1_MHDc} and \eqref{e:s2_MHDc} into
Equations \eqref{e:pho_p2}--\eqref{e:momx_jp2} and
\eqref{e:momz_jp2}--\eqref{e:indx_jp2}, and arrange terms to obtain
\begin{equation}\label{e:mat}
\mathbf{A}\cdot ( C_{\rm f1}, C_{\rm f2}, C_{\rm s2+}, C_{\rm
  s2-}, Z)^{T} = 0\,,
\end{equation}
where $\mathbf{A}$ is a $5\times5$ matrix whose components are given by
\begin{equation}
  \begin{split}\label{e:mat1}
    A_{11} & = 1 - \Msu^2 \chi_{\rm f1}/(\nu_{\rm f1}\sigma_{D, {\rm f1}})\,, \\
    A_{1j} & = 0\,,\\
    A_{15} & = -\sigma\,, \\
  \end{split}
\end{equation}
\begin{equation}\label{e:mat2}
  \begin{split}
    A_{21} & = 0\,, \\
    A_{2j} & = 1 - \Msd^2 \chi_{j}/(\alpha\nu_j \sigma_{D,j})\,, \\
    A_{25} & = -\sigma\,,
  \end{split}
\end{equation}
\begin{equation}\label{e:mat3}
  \begin{split}
    A_{31} & = \chi_{\rm f1}/(\nu_{\rm f1}\sigma_{D, {\rm f1}}^2) -
    \MAu^{-2}/\sigma_{D,{\rm f1}}\,, \\
    A_{3j} & = \chi_j/(\nu_j\sigma_{D,j}^2) -
    \alpha\MAd^{-2}/\sigma_{D,j}\,, \\
    A_{35} & = 1 + \MAu^{-2} - \alpha(1 + \MAd^{-2})\,,
  \end{split}
\end{equation}
\begin{equation}\label{e:mat4}
  \begin{split}
    A_{41} & = -\mathcal{B}\chi_{\rm f1}/(\nu_{\rm f1}\sigma_{D,{\rm
        f1}})\,, \\
    A_{4j} & = (1 + \betad^{-1})(1 - \Mmsd^2)\chi_j/(\alpha \nu_j
    \sigma_{D,j})\,, \\
    A_{45} & = (1 - \alpha^{-1})G\,,
  \end{split}
\end{equation}
\begin{equation}\label{e:mat5}
  \begin{split}
    A_{51} & = -\nu_{\rm f1}\MAu^{-1}/\sigma_{D,{\rm f1}} +
    \MAu^{-1}\Msu^2\chi_{\rm f1}/(\nu_{\rm f1}\sigma_{D,{\rm f1}})\,, \\
    A_{5j} & = \alpha^{3/2}\MAd^{-1}\nu_{j}/\sigma_{D,j} -
    \MAd^{-1}\Msd^2\chi_j/(\alpha^{1/2}\nu_j\sigma_{D,j})\,, \\
    A_{55} & = 0\,.
  \end{split}
\end{equation}
Here,
\begin{equation}\label{e:mat6}
  \mathcal{B} = (1 + \betau^{-1})(1 - \Mmsu^2) +
  \dfrac{\MAu^{-2}(1 - \alpha^{-2})}{(\sigma + \nu_{\rm
      f1})^2}\,,
\end{equation}
and the index $j$ in Equations \eqref{e:mat2}--\eqref{e:mat5} runs
from 2 to 4, with $j=2$, 3, and 4 corresponding to the downstream fast
($\rm{f2}$), forward-propagating slow ($\rm{s2+}$), and
backward-propagating slow ($\rm{s2-}$) modes, respectively.

\end{document}